\documentclass[tightenlines,eqsecnum,floats,aps,amsmath,amssymb%
,nofootinbib,prd,showpacs,twocolumn]{revtex4}

\usepackage{amsmath,amssymb,amsfonts}
\usepackage{graphicx}
\usepackage{enumerate}

\usepackage{LQC-symbols}

\usepackage{hyperref}

\begin{document}

\title{Physical evolution in Loop Quantum Cosmology: The example of vacuum
Bianchi I}
\author{Mercedes Mart\'{i}n-Benito${}^1$}
\email{merce.martin@iem.cfmac.csic.es}
\author{Guillermo A. Mena Marug\'{a}n${}^1$}
\email{mena@iem.cfmac.csic.es}
\author{Tomasz Pawlowski${}^1$}
\email{tomasz@iem.cfmac.csic.es}

\affiliation{${}^1$Instituto de Estructura de la
Materia, CSIC, Serrano 121, 28006 Madrid, Spain}

\begin{abstract}
We use the vacuum Bianchi I model as an example to
investigate the concept of physical evolution in Loop
Quantum Cosmology (LQC) in the absence of the massless
scalar field which has been used so far in the
literature as an internal time. In order to retrieve
the system dynamics when no such a suitable clock
field is present, we explore different constructions
of families of unitarily related partial observables.
These observables are parameterized, respectively, by:
$(i)$ one of the components of the densitized triad,
and $(ii)$ its conjugate momentum; each of them
playing the role of an evolution parameter. Exploiting
the properties of the  considered example, we
investigate in detail the domains of applicability of
each construction. In both cases the observables
possess a neat physical interpretation only in an
approximate sense. However, whereas in case $(i)$ such
interpretation is reasonably  accurate only for a
portion of the evolution of the universe, in case
$(ii)$ it remains so during all the evolution (at
least in the physically interesting cases). The
constructed families of observables are next used to
describe the evolution of the Bianchi I universe. The
performed analysis confirms the robustness of the
bounces, also in absence of matter fields, as well as
the preservation of the semiclassicality through them.
The concept of evolution studied here and the
presented construction of observables are applicable
to a wide class of models in LQC, including
quantizations of the Bianchi I model obtained with
other prescriptions for the improved dynamics.
\end{abstract}


\maketitle

\section{Introduction}

Loop Quantum Cosmology (LQC) \cite{lqc} is an area of
physics which deals with the quantization of symmetry
reduced gravitational systems by adopting similar
methods to those employed in Loop Quantum Gravity
\cite{lqg}.

Not surprisingly, the kinematical structure underlying
LQC was rigorously established for the first time for
the simplest of all the cosmological models, namely,
those describing a homogeneous, isotropic, and
spatially flat geometry \cite{abl}. The analysis of
the dynamics of such a kind of geometry in the
presence of a massless and minimally coupled scalar
field has shown that the big bang singularity is
resolved dynamically and is replaced by a quantum
bounce \cite{aps-let,aps-imp}. LQC has been further
extended, with diverse levels of rigor, to other
similar models with different topology \cite{top} or
nonvanishing cosmological constant \cite{lambda}, and
furthermore to more general settings such as
anisotropic systems \cite{chio,mmp,awe-b1}, or even to
inhomogeneous situations \cite{mgm-gowdy}. The
robustness of the singularity resolution features has
been confirmed within an exactly solvable version of
LQC \cite{acs,cs,b1-szulc}, and the mathematical
foundations of its elements have been discussed
\cite{kl-sadj,sadj}. In addition to the studies of the
genuine quantum theory, there exists an extensive
amount of work at the level of the effective classical
dynamics \cite{sv-eff,victor}, which provides
important insights into the properties of the quantum
geometry in cosmological scenarios
\cite{eff-rep,Geff}.

So far, all the analyses of the dynamics carried
out in the genuine quantum theory have employed
partial observables parameterized by an emergent time.
In most of the models considered in the literature
\cite{aps-let,aps-imp,top,lambda,chio,acs,b1-szulc,awe-b1},
the role of time was played by a massless scalar
field, which (unlike the geometry degrees of freedom)
was quantized adopting a standard Schroedinger-like
representation. However, in many cases (vacuum
homogeneous universes, black hole interiors) such
possibility is not at hand. In this article, we
overcome this difficulty by constructing and studying
the properties of various families of unitarily
related observables parameterized by the geometry
degrees of freedom. As an appropriate test bed for our
analysis, we study a model of a vacuum homogeneous
universe of the Bianchi I type.

The Bianchi I system itself has been extensively
analyzed in the literature. After a preliminary
analysis of the kinematics \cite{boj-bianchi}, an
attempt to quantize the model was made in Ref.
\cite{chio}, and the corresponding effective dynamics
was studied in Ref. \cite{chi2}. There, however, a
homogeneous massless scalar field was coupled as
matter content and identified as an internal time.

More recently, the system in vacuo has been thoroughly
quantized \cite{mmp}. In that work, the studies were
restricted to the model with the compact three-torus
($T^3$) topology, in order to provide a basis for
further studies of a much more general system, namely
the inhomogeneous Gowdy $T^3$ model with linear
polarization \cite{mgm-gowdy}. In Ref. \cite{mmp} both
the physical Hilbert space and a complete algebra of
observables were constructed; nonetheless the dynamics
of the system was not investigated.

Let us comment that the extension of a quantization
prescription known as improved dynamics from the
isotropic to the anisotropic setting in homogenous LQC
has given rise to the proposal of two concurrent
prescriptions for Bianchi I. In this article,
following Ref. \cite{mmp}, we use the prescription
proposed in Ref. \cite{chio}. For noncompact models
and just in the form presented in that paper, that
prescription is not invariant under certain
transformations \cite{adv}, but for compact models,
including the one considered in Ref. \cite{mmp}, the
prescription is free from this drawback. The desired
invariance is recovered in noncompact situations with
an alternative prescription, which was recently
developed and motivated in Ref. \cite{awe-b1}
appealing to the hypothesized relationship between the
degrees of freedom of LQC and full LQG. For our
discussion, nonetheless, the following reasons explain
our use of the former of these prescriptions. $(a)$
The goal of our work is a methodological development
of a formalism of unitary evolution, defining it in a
precise manner without introducing an additional
matter field as a clock. For that purpose, the model
of Ref. \cite{mmp} is a better candidate, as the
structure of the Hilbert space and the properties of
the states are known in detail and have been presented
in the literature, something which is necessary to
establish the link between physical intuition and the
exact mathematical implementation of our method. $(b)$
The specific properties of the model of Ref.
\cite{mmp} allow to study the limitations of
applicability of some of the constructed methods (see
in particular Fig. \ref{fig:v1-v2-comp} and the
related discussion in Subsec. \ref{sec:v-results}),
limitations which would have been missed had one used
the model of Ref. \cite{awe-b1}. $(c)$ Our
construction, although applied here to a concrete
model, is intended to be reasonably universal. In
particular, it is almost directly applicable to other
possible prescriptions for the quantization of the
Bianchi I model, including that of Ref. \cite{awe-b1}
(see the discussion in Sec. \ref{sec:results}).

Employing the knowledge accumulated in previous works
about the properties of the selected model, and
the explicit and complete quantization chosen for it,
we will analyze here two possible constructions of
families of unitarily related observables, which are
respectively parameterized by: $(i)$ a coefficient of
the densitized triad, and $(ii)$ the variable
conjugate to it. For the sake of clarity in the
presentation, as well as for comparing the predictions
of geometrodynamics and LQC, we will apply the
construction to both the LQC version of the model and
its Wheeler-DeWitt (WDW) counterpart. We will see
that, whereas in a WDW-type quantization our
construction with respect to both of the specified
parameters provides observables with a precise
physical interpretation, in the loop quantization
scheme this property is achieved only in an
approximate sense. Furthermore, the accuracy of the
interpretation depends only on properties of the state
in case $(ii)$, while for $(i)$ the dependence is also
on the evolution epoch.

The paper is organized as follows. Sec.
\ref{sec:framework} is a review of the LQC
quantization of Ref. \cite{mmp}. In Sec.
\ref{sec:wdw-analog} we perform the WDW quantization
of the model and develop an interpretation for the
evolution of observables for the two considered
choices of internal time. In Sec. \ref{sec:wdw-limit}
we show that the LQC states converge in a certain
limit to a combination of WDW states. The evolution
picture in the LQC theory is analyzed in Sec.
\ref{sec:v-evo} and Sec. \ref{sec:lqc-b-rep} for the
choices of internal time $(i)$ and $(ii)$,
respectively. In Sec. \ref{sec:results} we discuss the
main results of this work. Besides, two appendices are
added. One of them extends the discussion on the WDW
regime to take into account the union of different
sectors. Finally, a brief description of the effective
classical dynamics, associated with the genuine LQC
dynamics, is given in Appendix \ref{app:eff}.

\section{The framework}\label{sec:framework}

In this section we briefly review the basic features
of the polymerically quantized vacuum Bianchi I model.
It is a summary of the main results obtained in Ref.
\cite{mmp}, which the reader can consult for further
details.

\subsection{Classical theory}

The Bianchi I model represents spatially flat and
homogeneous spacetimes. We will consider the case of a
compact topology: that of a three-torus. The spacetime
metric can be written in the form \cite{chi2}
\begin{align}\label{metric}
ds^2&= -N^2 dt^2+\frac{|p_1p_2p_3|}{4\pi^2}\sum_{i=1}^3
\frac{(dx^i)^2}{p_i^2},
\end{align}
where $\{dx^i\}$ is the fiducial co-triad, $N$ is the
lapse function, and $p_i/(4\pi^2)$ are the nontrivial
components of the densitized triad in a diagonal
gauge. The corresponding components of the Ashtekar
connection are $c^i/(2\pi)$, such that
$\{c^i,p_j\}=8\pi G\gamma\delta^i_j$. Here $G$ is the
Newton constant and $\gamma$ is the Immirzi parameter
(that we assume positive for simplicity). Owing to
homogeneity, the only constraint present in the model
is the Hamiltonian one, given by \cite{chi2}
\begin{align}\label{H}
C_\text{BI}&=-\frac{2}{\gamma^2}\frac{c^1p_1c^2p_2
+c^1p_1c^3p_3+c^2p_2c^3p_3}{V}=0,
\end{align}
where $V=\sqrt{|p_1p_2p_3|}$ is the spacetime volume.
Classically, any of the triad coefficients $p_i$ is a
monotonous function of the time coordinate $t$
\cite{chi2}. Moreover, the spacetime presents a
curvature singularity at initial time $t=0$, at which
the universe stretches as an infinitely long line.

\subsection{Loop Quantum Cosmology
kinematics}\label{sec:fr-kin}

In LQC the basic configuration variables are
holono\-mies of connections. The holonomy along an
edge of oriented coordinate length $2\pi\mu_i$ in the
direction $i$ is defined as $h_i^{\mu_i}(c^i)=e^{
\mu_{i}c^{i}\tau_{i}}$, where $\tau_i$ are the $SU(2)$
generators proportional to the Pauli matrices, such
that $[\tau_i,\tau_j]=\epsilon_{ijk}\tau^k$. The
configuration algebra $\text{Cyl}_\text{S}$ is the
algebra of almost periodic functions of $c^i$, which
is generated by the matrix elements of the holonomies
$\mathcal N_{\mu_i}(c^i)=e^{\frac{i}{2}\mu_{i}c^{i}}$.
In the momentum representation, the states defined by
these matrix elements are denoted by $|\mu_i\rangle$.
The completion of the algebra $\text{Cyl}_\text{S}$
with respect to the discrete inner product
$\langle\mu_i|\mu_i^\prime\rangle=\delta_{\mu_i
\mu_i^\prime}$ for each fiducial direction provides
the kinematical Hilbert space $\mathcal
H_{\text{Kin}}=\otimes_i \mathcal H_{\text{Kin}}^i$.
The elementary operators are the operators $\hat p_i$
associated with fluxes, which are diagonal on the
basis states $|\mu_i\rangle$ of $\mathcal
H_{\text{Kin}}^i$, and $\hat{\mathcal
N}_{\mu_i^\prime}$, whose action shifts the label
$\mu_i$ of the considered basis by $\mu_i^\prime$.

In order to express $C_\text{BI}$ in terms of
holonomies, instead of connections components, one
introduces the curvature tensor associated with the
Ashtekar connection and defines it as in gauge lattice
theories. With this aim, one considers a loop of
holonomies and shrinks its area to its minimum value,
which is nonzero owing to the discreteness of the area
spectrum in LQG. To determine the minimum
$2\pi\bar\mu_i$ for the coordinated length of the
holonomy we assume (as in Ref. \cite{chio}) square
loops of fiducial area $4\pi^2\bar{\mu}_i^2$ with
minimum physical area equal to
$\Delta=2\sqrt{3}\pi\gamma l_{\text{Pl}}^2$, with
$l_{\text{Pl}}=\sqrt{G\hbar}$ being the Planck length.
This implies that $\bar{\mu}_i^2|p_i|=\Delta$ in the
sense of operators \cite{F3}. On the other hand,  one
appeals to the so called ``Thiemann's trick'' to
define a well-defined inverse volume operator
$\widehat{[1/V]}$. As a final result one obtains a
quantum Hamiltonian constraint $\widehat C_\text{BI}$
expressed in terms of our basic operators and which is
well-defined in the domain $\text{Cyl}_\text{S}$
\cite{chio,mmp}.

Owing to the relation $\bar{\mu}_i^2|p_i|=\Delta$, the
shift produced by the operator $\hat{\mathcal
N}_{\pm\bar\mu_i}$ on the basis states $|\mu_i\rangle$
is not constant. One can nevertheless relabel these
states with an affine parameter
\begin{equation}\label{eq:v-def}
v_i(\mu_i) = 2^{\frac{3}{2}} 3^{-\frac{5}{4}}\sgn(\mu_i)
|\mu_i|^{\frac{3}{2}},
\end{equation}
to make the shift uniform \cite{chio,mmp}. The action
of the basic operators in the relabeled states is
$\hat{\mathcal
N}_{\pm\bar\mu_i}|v_i\rangle=|v_i\pm1\rangle$ and
$\hat p_i|v_i\rangle=3^{1/3}\Delta\,\text{sgn}(v_i)
|v_i|^{2/3}|v_i\rangle$.

We have symmetrized the resulting Hamiltonian
constraint $\widehat{C}_{\text{BI}}$ in such a way
that it annihilates the subspace spanned by the
``zero-volume'' states, i.e. the states
$\otimes_i|v_i\rangle$ with any $v_i$ equal to zero,
and leaves invariant its orthogonal complement. For
the nontrivial solutions to the constraint, we can
thus restrict our study to that complement. Owing to
this restriction the singularity is resolved at the
kinematical level \cite{mmp}.

It is convenient to densitize the Hamiltonian
constraint to find its solutions. The densitized
Hamiltonian constraint is given by the operator
\begin{equation}\label{eq:Qdens}
\widehat{{\cal C}}_{\text{BI}}=
\widehat{\left[\frac1{V}\right]}^{-\frac1{2}}
\widehat{C}_{\text{BI}}
\widehat{\left[\frac{1}{V}\right]}^{-\frac1{2}},
\end{equation}
which is well-defined in the considered domain (linear
span of tensor products of states $|v_i\rangle$ such
that none of the $v_i$'s vanishes). The bijection
$(\tilde\psi|=(\psi|\widehat{[1/V]}{}^{1/2}$ in the
dual of that domain relates the nontrivial solutions
$(\psi|$ of $\widehat{C}_{\text{BI}}$ to the solutions
$(\tilde\psi|$ of its densitized version
$\widehat{{\cal C}}_{\text{BI}}$. This operator turns
out to be given by
\begin{equation}\label{C}
\widehat{{\cal C}}_{\text{BI}}=-\frac{2}{\gamma^2}
\bigg[\widehat{\Theta}_1
\widehat{\Theta}_2+\widehat{\Theta}_1\widehat{\Theta}_3+
\widehat{\Theta}_2\widehat{\Theta}_3\bigg],
\end{equation}
where the operator $\widehat{\Theta}_i$ is symmetric
in the domain spanned by the basis states
$|v_i\rangle$ ($v_i\neq0$). Its action is:
\begin{equation}\label{acttheta}
\widehat{\Theta}_i|v_i\rangle=-i\frac{\Delta}{2\sqrt{3}}
\big[f_+(v_i)|v_i+2\rangle-f_-(v_i)|v_i-2\rangle\big],
\end{equation}
where
\begin{align}\label{f}
&f_\pm(v_i)=g(v_i\pm2)s_\pm(v_i)g(v_i), \\\label{s}
&s_\pm(v_i)=\text{sgn}(v_i\pm2)+\text{sgn}(v_i),\\\label{g}
&g(v_i)=
\begin{cases}
\left|\left|1+\frac1{v_i}\right|^{\frac1{3}}
-\left|1-\frac1{v_i}\right| ^{\frac1{3}}
\right|^{-\frac1{2}} & {\text{if}} \quad v_i\neq 0,\\
0 & {\text{if}} \quad v_i=0.\\
\end{cases}
\end{align}

Let us note that the operator $\widehat{\Theta}_i$
leaves invariant the Hilbert subspaces $\mathcal
H_{\varepsilon_i}^{\pm}$ defined as the Cauchy
completions with respect to the discrete inner product
of the spaces
\begin{equation}
\text{Cyl}_{\varepsilon_i}^{\pm}=\text{span}\{|v_i\rangle;
v_i\in\mathcal L_{\varepsilon_i}^\pm\},
\end{equation}
where $\mathcal L_{\varepsilon_i}^\pm$ are the
semilattices of step two
\begin{equation}
\mathcal{L}_{\varepsilon_i}^\pm=
\{\pm(\varepsilon_i+2n),n=0,1,2,...\},
\quad\varepsilon_i\in(0,2].
\end{equation}
Therefore the constraint $\widehat{{\cal
C}}_{\text{BI}}$ superselects the kinematical Hilbert
space in different separable sectors. If we choose,
for instance, a positive orientation for the triad in
the three fiducial directions, we can restrict our
study to the kinematical Hilbert space $\mathcal
H_{\vec\varepsilon}^+=\otimes_i\mathcal
H_{\varepsilon_i}^+$, with
$\vec\varepsilon=(\varepsilon_1,\varepsilon_2,
\varepsilon_3)$.

\subsection{Physical Hilbert space}
\label{sec:fr-Hph}

In order to obtain the physical Hilbert space, it is
enough to analyze the spectral properties of the
operator $\widehat{\Theta}_i$. It can be proven
essentially self-adjoint with domain the dense set
$\text{Cyl}_{\varepsilon_i}^{+}$. In consequence,
$\widehat{{\cal C}}_{\text{BI}}$ is essentially
self-adjoint in the invariant domain
$\text{Cyl}^+_{\vec\varepsilon}=\otimes_i
\text{Cyl}_{\varepsilon_i}^+$.
This property allows us to apply the group averaging
method \cite{gave,gave2} to determine the physical
Hilbert space $\mathcal
H^\text{Phy}_{\vec\varepsilon}$.

In addition, the spectrum of $\widehat{\Theta}_i$ is
absolutely continuous, coincides with the real line
and is nondegenerate. Furthermore, the coefficients
$e^{\varepsilon_i}_{\omega_i}(2n+\varepsilon_i)$
($n\in \mathbb{N}^+$) of its generalized
eigenfunctions with eigenvalue $\omega_i$ turn out to
be determined just by the initial data
$e^{\varepsilon_i}_{\omega_i}(\varepsilon_i)$ [see
Ref. \cite{mmp} for the explicit expression -Eq.
(45)-].

These eigenfunctions are formed by two components, one
of them with support in the semilattice of step four
\[^{(4)}{\mathcal L}_{\varepsilon_i}^+= \{(\varepsilon_i+4n),
n=0,1,2,...\}\] and the other in the displaced
semilattice
\[^{(4)}{\mathcal L}_{\varepsilon_i+2}^+=
\{(\varepsilon_i+2+4n),
n=0,1,2,...\}.\] Both of these components are
generalized eigenfunctions with eigenvalue
$\omega_i^2$ of the operator $\widehat{\Theta}_i^2$,
which is proportional to the gravitational part of the
densitized constraint in the isotropic case. These
components have a relative phase of $\pm\pi/2$ and,
hence, $e_{\omega_i}^{\varepsilon_i}(v_i)$ oscillates
rapidly when $v_i$ varies in the semilattice
${\mathcal L}_{\varepsilon_i}^+$.

Applying the group averaging procedure, we obtain the
following form for the wave function of the physical
states
\begin{align}\label{eq:fdec}
\Phi(\vec{v}) &= \int_{\re^2} \rd\omega_2
\rd\omega_3\, \tilde\Phi(\omega_2,\omega_3)\,
e^{\varepsilon_1}_{\omega_1(\omega_2,\omega_3)}(v_1)
\nonumber\\
&\times e^{\varepsilon_2}_{\omega_2} (v_2)
e^{\varepsilon_3}_{\omega_3}(v_3) ,
\end{align}
with
\begin{equation}\label{om-func}
\omega_1(\omega_2,\omega_3)=
-\frac{\omega_2\omega_3}{\omega_2+\omega_3}.
\end{equation}
Here $\tilde\Phi(\omega_2,\omega_3)$ belongs to the
physical Hilbert space, which turns out to be
\begin{equation}\label{physpa}
\mathcal H^\text{Phy}_{\vec\varepsilon}= L^2\left({\re^2},
|\omega_2+\omega_3|\rd\omega_2 \rd\omega_3\right).
\end{equation}

In what follows, unless otherwise specified,
$\omega_1$ will be the function
$\omega_1(\omega_2,\omega_3)$ given in Eq.
\eqref{om-func}.

\section{Wheeler-DeWitt analog}\label{sec:wdw-analog}

In this section we will study the quantization of the
vacuum Bianchi I model in the WDW approach formulated
in terms of the connection coefficients, which will be
treated in a standard (non-polymeric) way.

\subsection{Kinematics and scalar constraint}

Like in the loop quantization, we will work in the
triad representation. As kinematical Hilbert space of
the WDW quantization we take
$\ub{\Hil}_{\kin}=\otimes_i\ub{\Hil}_{\kin}^i$ with
$\ub{\Hil}_{\kin}^i=L^2(\mathbb{R},dv_i)$ \cite{F1}.
The measure is the usual Lebesgue measure, and not the
discrete one. In this representation the operator
$\hat p_i$ acts by multiplication by the factor
$p_i=3^{1/3}\Delta\,\text{sgn}(v_i) |v_i|^{2/3}$, just
as in the loop quantization. On the other hand, we
promote the connection coefficients to derivative
operators,
\begin{equation}
\hat c^i=i3^{1/6}2|v_i|^{1/3}{\partial}_{v_i},
\end{equation}
to preserve the Dirac rule $[\hat c {}^i,
\hat{p}_j]=i\hbar\widehat{\{c^i,p_j\}}$.

We denote by $\widehat{\underline\Theta}_i$ the
counterpart of the classical quantity $c^ip_i$ in the
WDW quantum theory [defined on the Schwartz space
$\mathcal{S}(\re)$]. We choose for
$\widehat{\underline\Theta}_i$ the symmetric factor
ordering which is analog to that used in the loop
quantization, to simplify the comparison with it:
\begin{equation}
\widehat{\underline\Theta}_i=i \sqrt{3}
\Delta\sqrt{|v_i|}[\text{sgn}(v_i)\partial_
{v_i}+\partial_
{v_i}\text{sgn}(v_i)]\sqrt{|v_i|}.
\end{equation}
It is well defined in the distributional sense and can
be rewritten in the simpler form
$\widehat{\underline\Theta}_i=i
\sqrt{3}\Delta(1+2v_i\partial_{v_i})$, where we have
disregarded the noncontributing term
$|v_i|\delta(v_i)$. Then, the WDW quantum counterpart
of the classical densitized Hamiltonian constraint can
be constructed exclusively from the operator
$\widehat{\underline\Theta}_i$ in the following manner
\begin{equation}\label{CWDW}
\widehat{{\underline{\cal C}}}_{\text{BI}} =
-\frac{2}{\gamma^2}\bigg[\widehat{\underline\Theta}_1
\widehat{\underline\Theta}_2
+ \widehat{\underline\Theta}_1\widehat{\underline\Theta}_3
+ \widehat{\underline\Theta}_2\widehat{\underline\Theta}_3
\bigg].
\end{equation}

The operator $\widehat{\underline\Theta}_i$ is
essentially self-adjoint on $\ub{\Hil}_{\kin}^i$.
Furthermore, its restrictions to each of the subspaces
$\ub{\Hil}_{\kin}^{i,\pm}=L^2(\mathbb{R}^\pm,dv_i)$
are essentially self-adjoint as well. On each of these
subspaces, the spectrum of
$\widehat{\underline\Theta}_i$ is absolutely
continuous, coincides with the real line and is
nondegenerate, as happens to be the case with its
analog $\widehat{\Theta}_i$ in the loop quantization.
Moreover, its generalized eigenfunctions, with
generalized eigenvalue $\omega_i$, have the form
\begin{equation}\label{eq:WDW-eig}
\underline{e}_{\omega_i}(v_i)= \frac{1}{\sqrt{2\pi\alpha
|v_i|}}\exp\left({-i\omega_i\frac{\ln{|v_i|}}{\alpha}}\right),
\end{equation}
where $\alpha=2\sqrt{3}\Delta=12\pi\gamma
l_{\text{Pl}}^2$. They provide an orthonormal basis
for $\ub{\Hil}_{\kin}^{i,\pm}$. In analogy with the
procedure followed in the loop quantization, we will
restrict the study to
$\ub{\Hil}_{\kin}^{+}=\otimes_i\ub{\Hil}_{\kin}^{i,+}$.

\subsection{Physical Hilbert space and observables}
\label{sec:WDWphys-obs}

In order to obtain the physical Hilbert space, we
apply the group averaging method like in the loop
quantization. Obviously, the physical Hilbert space
obtained is
\begin{equation}\label{physpaWDW}
\underline{\mathcal H}^\text{Phy}= L^2\left(\mathbb{R}^2,
|\omega_2+\omega_3|\rd\omega_2 \rd\omega_3\right).
\end{equation}
The wave function of the physical states has the
following form:
\begin{align}\label{phystawdw}
{\underline\Phi}(\vec{v}) &= \int_{\mathbb{R}^2}
\rd\omega_2 \rd\omega_3\,
{\underline{\tilde\Phi}}(\omega_2,\omega_3)\, \underline
e_{\omega_1}(v_1)
\nonumber\\
&\times\underline e_{\omega_2} (v_2)\underline e_{\omega_3}(v_3),
\end{align}
with $ {\underline{\tilde\Phi}}(\omega_2,\omega_3)
\in\underline{\mathcal
H}^\text{Phy}$.

As in the LQC model \cite{mmp}, the operators
$\widehat{\ub{\Theta}}_a$ (here $a=2,3$), which
multiply the wave function by $\omega_a$, together
with $-i|\omega_2+\omega_3|^{-1/2}\partial_{\omega_a}
|\omega_2+\omega_3|^{1/2}$, provide a complete set of
observables, which are essentially self-adjoint
operators on the domain $\mathcal
S(\re^2)\subset\ub{\Hil}^{\phy}$. Nonetheless, while
the first pair are Dirac observables which correspond
to classical constants of motion, the second one are
not. As a consequence, this set turns out not to be
adequate to introduce a nontrivial concept of
evolution. In the next subsection, we will construct
another pair of Dirac observables that will complete
the set formed by $\widehat{\ub{\Theta}}_a$ ($a=2,3$),
and in terms of which we will be able to develop an
interpretation for the notion of evolution.

\subsection{The evolution}\label{sec:WDWphys-evo}

In any gravitational system, as the one considered
here, the Hamiltonian is constrained to vanish.
Therefore, there is no well defined notion of
evolution in the model. Nonetheless, one can try to
select, in the configuration space, a coordinate $t$
(an internal time) and define a map between Hilbert
spaces $\ub{\Hil}_t $, where $t$ takes values in a
certain set $\mathcal{U}$ and $\ub{\Hil}_t$ is the
space of ``initial data'' given by the restriction of
the wave function to the surface $t=\const$. If there
exists a unitary transformation $\ub
P_t:\ub{\Hil}^{\phy}\to\ub{\Hil}_t$, then each
$t$-slice contains all the information needed to
determine the physical state (i.e. the system is
closed). Furthermore, if, in addition, the
``identity'' map between the spaces $\ub{\Hil}_t$
(given by the trivial identification of data at
different times) is also unitary, one can define a
unitary evolution in $\ub{\Hil}_t$. This is achieved
by composing the inverse transformation $\ub
P_t^{-1}$, a transformation in the family $\ub
P_t$ (for a different value of $t$), and the above
identification of data.

In our case, since classically any of the three triad
components is monotonous along all the dynamical
trajectories, one can select one of the $v_i$'s as an
internal time. Taking into account that in the
description of the physical states \eqref{phystawdw}
we have already eliminated $\omega_1$ in terms of
$\omega_2$ and $\omega_3$, it is then most natural to
choose $t:=v_1$.

We can easily introduce the initial data spaces
labeled by $v_1=\const$ (which we will also call
``slice'' spaces) via the transformation
\begin{subequations}\label{eq:wdw-sl-def}\begin{align}
\ub{\Phi}_{v_1}(v_2,v_3) &= \int_{\re^2} \rd\omega_2\rd\omega_3
\ub{\tilde{\Phi}}_{v_1}(\omega_2,\omega_3)
\ub{e}_{\omega_2} (v_2) \ub{e}_{\omega_3}(v_3), \notag \\
\begin{split}
\ub{\tilde{\Phi}}_{v_1}(\omega_2,\omega_3) &:=
\ub{P}_{v_1}\ub{\tilde{\Phi}}(\omega_2,\omega_3) \\
&:= \ub{\tilde{\Phi}}(\omega_2,\omega_3)
\ub{e}_{\omega_1}(v_1),
\end{split}
\tag{\ref{eq:wdw-sl-def}}
\end{align}\end{subequations}
 with $\ub{\tilde{\Phi}}_{v_1}(\omega_2,\omega_3)$ belonging to
\begin{equation}\label{eq:wdw-sl-H-def}\begin{split}
\ub{\Hil}_{v_1} &= L^2(\re^2,|\omega_2+\omega_3|
|\ub{e}_{\omega_1}(v_1)|^{-2}
\rd\omega_2\rd\omega_3) \\
&= L^2(\re^2,2\pi\alpha v_1|\omega_2+\omega_3|\rd\omega_2\rd\omega_3).
\end{split}\end{equation}
Since $\ub{e}_{\omega_1}(v_1)$, given in Eq.
\eqref{eq:WDW-eig}, never vanishes, all the  slice
spaces $\ub{\Hil}_{v_1}$ are unitarily related to
$\ubHilp$.

Introducing an additional rescaling one can define the
alternate transformation
\begin{subequations}\label{eq:wdw-sl-def-mod}\begin{align}
\ub{\Phi}_{v_1}(v_2,v_3) &= \int_{\re^2}
\frac{\rd\omega_2\rd\omega_3}{\sqrt{2\pi\alpha v_1}}
\ub{\tilde{\Phi}}'_{v_1}(\omega_2,\omega_3)
\ub{e}_{\omega_2} (v_2) \ub{e}_{\omega_3}(v_3), \notag \\
\begin{split}
\ub{\tilde{\Phi}}'_{v_1}(\omega_2,\omega_3) &:=
\ub{P}'_{v_1}\ub{\tilde{\Phi}}(\omega_2,\omega_3) \\
&:= \ub{\tilde{\Phi}}(\omega_2,\omega_3)
\sqrt{2\pi\alpha v_1}\ub{e}_{\omega_1}(v_1) \in \ub{\Hil}'_{v_1},
\end{split}
\tag{\ref{eq:wdw-sl-def-mod}}
\end{align}\end{subequations}
where
\begin{subequations}\begin{align}
\ub{P}'_{v_1} &: \ub{\Hil}^{\phy}\to\ub{\Hil}'_{v_1} \\
\ub{\Hil}'_{v_1} &=
L^2(\re^2,|\omega_2+\omega_3|\rd\omega_2\rd\omega_3).
\end{align}\end{subequations}
Since $\ub{P}'_{v_1}$ is unitary and the spaces
$\ub{\Hil}'_{v_1}$ coincide (so that the corresponding
map identifying states at different values of $v_1$ is
unitary as well), it is clear that $\ub{P}'_{v_1}$
leads to a unitary evolution for the system.

A more elaborated manner of providing a concept of
evolution is by introducing a family of \emph{partial
observables} \cite{obs,bmgmm}, which are to be related
via unitary transformations. A natural construction of
partial observables from the kinematical ones is
available if there exists an internal time which
provides a unitary map between initial data spaces. On
the formal level, one can do this through the group
averaging procedure. Alternatively, if there exists a
decomposition $\ubHilk=\ub{\Hil}^t\otimes\ub{\Hil}'$
(where $\ub{\Hil}^t$ is some Hilbert space of
functions depending on $t$ only)\cite{F2} and there is
an operator $\hat{O}':\ub{\Hil}'\to\ub{\Hil}'$, one
can build $\hat{O}_t$ (which measures the
corresponding quantity ``at a given time $t$'') as an
operator whose action is defined through the following
sequence of operations:
\begin{enumerate}[(i)]
\item Take an initial data slice $\psi_t\in\ub{\Hil}_t$
corresponding to the value $t$ of an emergent time.

In many cases the spaces $\ub{\Hil}_t$ differ from
$\ub{\Hil}'$. Therefore one has to define a
transformation $\ub{\Hil}_t\to\ub{\Hil}'$ \cite{AltH}.
\item Then act with the ``kinematical'' observable $\hat{O}'$
corresponding to the measured quantity,
and \label{it:obs-pkin}
\item use the relation between physical and slice Hilbert
spaces to find the element of $\ub{\Hil}^{\phy}$
corresponding to the result of \eqref{it:obs-pkin}.
\end{enumerate}

In the model under study, the above method can be
applied for example to the kinematical observable
$\ln(\hat{v}_a)$ (where $a=2,3$), which acts on
elements of $\ub{\Hil}_{\kin}^{a,+}$ as a
multiplication operator. As the internal time
coordinate we still choose $v_1$, so that
$\ub{\Hil}_t:=\ub{\Hil}_{v_1}$, given by Eq.
\eqref{eq:wdw-sl-H-def}. The space $\ub{\Hil}'$ is the
product
\begin{equation}\label{eq:wdw-Hprime}
\ub{\Hil}' := \ub{\Hil}_{\kin}^{2,+}\otimes
\ub{\Hil}_{\kin}^{3,+} = L^2((\re^+)^2,\rd v_2 \rd
v_3)
\end{equation}

Since $\ub{\Hil}_{v_1}$ and $\ub{\Hil}'$ are
different, we introduce  a unitary transformation
$\ub{\Hil}_{v_1}\to\ub{\Hil}'$. In the corresponding
$\omega_a$-representations, it is given by the map
\begin{subequations}\label{eq:wdw-h'tr}\begin{align}
\tilde{\ub{\Phi}}_{v_1}(\omega_2,\omega_3)\, \mapsto
&\,\tilde{\ub{\boldsymbol{\chi}}}_{v_1}
(\omega_2,\omega_3) \tag{\ref{eq:wdw-h'tr}}\\
& := \sqrt{2\pi\alpha
v_1}|\omega_2+\omega_3|^{\frac{1}{2}}
\tilde{\ub{\Phi}}_{v_1}(\omega_2,\omega_3).\notag
\end{align}\end{subequations}
In the $v_a$-representation, this transformation is
simply the map $\ub{\Phi}_{v_1}(v_2,v_3)\to
\ub{\boldsymbol{\chi}}_{v_1}(v_2,v_3)$, with
$\ub{\Phi}_{v_1}(v_2,v_3)$ given in the first line of
Eq. \eqref{eq:wdw-sl-def} and
\begin{equation}\label{eq:aux-chi-wdw}
\ub{\boldsymbol{\chi}}_{v_1}(v_2,v_3) =
\int_{\re^2}\rd\omega_2\rd\omega_3
\tilde{\ub{\boldsymbol{\chi}}}_{v_1}(\omega_2,\omega_3)
\ub{e}_{\omega_2}(v_2) \ub{e}_{\omega_3}(v_3).
\end{equation}

Employing the transformations between the introduced
Hilbert spaces, we finally obtain a family of
observables $\ln(\hat{v}_a)_{v_1}$ acting on the
physical Hilbert space, interpretable as ``the value
of $\ln(v_a)$ at the fixed time $v_1$''. The operators
are well defined on the Schwartz space
$\mathcal{S}(\re^2)$. Their action in this domain,
which can be deduced taking into account that they act
by multiplication on the corresponding kinematical
states $\ub{\boldsymbol{\chi}}_{v_1}(v_2,v_3)$, turns
out to be
\begin{subequations}\label{obs}\begin{align}
&[\ln(\hat{v}_a)_{v_1}\ub{\tilde{\Phi}}](\omega_2,\omega_3)
= \frac{-i\alpha} {\ub{e}_{\omega_1}(v_1)}
|\omega_2+\omega_3|^{-\frac{1}{2}} \notag \\
&\hphantom{,}\times \partial_{\omega_a}\left[
|\omega_2+\omega_3|^{\frac{1}{2}}
\ub{\tilde{\Phi}}(\omega_2,\omega_3)
\ub{e}_{\omega_1}(v_1)\right] \tag{\ref{obs}}.
\end{align}\end{subequations}

The observables $\ln(\hat{v}_a)_{v_1}$, together with
the constants of motion $\widehat{\ub{\Theta}}_a
|_{v_1}:={\widehat{\ub{\Theta}}_a}$, form a complete
set of Dirac observables. Whereas within each family
(corresponding to $a=2,3$ respectively) the
observables $\widehat{\ub{\Theta}}_a |_{v_1}$ do not
change with $v_1$, the observables
$\ln(\hat{v}_a)_{v_1}$ do not coincide, and are
related at different times $v_1$ and $v_1^\star$ via
an operator $\ub{\widehat
Q}_{v_1,v_1^\star}:\underline{\mathcal
H}^\text{Phy}\rightarrow\underline{\mathcal
H}^\text{Phy}$ such that
\begin{equation}
[\ub{\widehat Q}_{v_1,v_1^\star}\underline{\tilde\Phi}]
(\omega_2,\omega_3)
= \sqrt{\frac{v_1}{v_1^\star}}
\frac{\underline e_{\omega_1}(v_1)}
{\underline e_{\omega_1}(v_1^\star)}
\underline{\tilde\Phi}(\omega_2,\omega_3).
\end{equation}
The form of the eigenfunction \eqref{eq:WDW-eig}
implies immediately, that these operators are both
invertible ($\ub{\widehat
Q}_{v_1,v_1^\star}^{-1}=\ub{\widehat
Q}_{v_1^\star,v_1}$) and unitary on
$\ub{\Hil}^{\phy}$. Therefore, the relation between
observables at different times,
\begin{equation}
\ln({\hat{v}}_a)_{v_1^\star}
= \ub{\widehat Q}_{v_1,v_1^\star}\,\ln(\hat{ v}_a)_{v_1}
\,\ub{\widehat Q}_{v_1^\star,v_1},
\end{equation}
is unitary.

As a consequence, the families of observables
$\ln(\hat{v}_a)_{v_1}$ define on the physical Hilbert
space $\ub{\Hil}^{\phy}$ a unitary evolution that is
local in the emergent time $v_1$. In contrast, as we
will see in Subsec. \ref{sec:pre-v-obs}, the direct
application of the above construction in the loop
quantization does not lead to a unitary evolution
since, in that case, the analogs of $\ub{\widehat
Q}_{v_1,v_1^\star}$ fail to be unitary operators.
Nonetheless (as we will see in Subsec.
\ref{sec:b-obs}) families of unitarily related
observables can be defined once we use, instead of
$v_1$, its conjugate momentum, denoted by $b_1$, which
provides a suitable emergent time in the loop
quantization.

To compare the dynamics predicted by the constructed
families of observables with the classical dynamics,
let us calculate the expectation values on some class
of states which are semiclassical at late times,
namely Gaussian states peaked around large values of
$\omega_2^\star$ and $\omega_3^\star$:
\begin{equation}\label{eq:wdw-gauss}
\ub{\tilde{\Phi}}(\omega_2,\omega_3) =
\frac{K}{\sqrt{|\omega_2+\omega_3|}}\prod_{a=2}^{3}
e^{-\frac{(\omega_a-\omega_a^\star)^2}{2\sigma_a^2}}
e^{i\beta^a\omega_a},
\end{equation}
where $K$ is a normalization factor such that
$\|\ub{\tilde{\Phi}}\|=1$, and the factor
$|\omega_2+\omega_3|^{-1/2}$ compensates the
nontrivial factor in the measure of the physical
Hilbert space \eqref{physpaWDW}.

For a general state $\ub{\tilde{\Phi}}$, using
directly the explicit form of the observables
\eqref{obs} and integrating the inner product, we find
that
\begin{equation}\label{eq:wdw-traj-shape}
\langle \ub{\tilde{\Phi}} | \ln(\hat{v}_a)_{v_1}
\ub{\tilde{\Phi}} \rangle
= A_a \ln v_1 + B_a ,
\end{equation}
where the constant coefficients $A_a$ and $B_a$ are
\begin{subequations}\label{eq:traj-dir}\begin{align}
\label{eq:traj-dir-A}
A_a &= \| \omega_1({\omega}_2,{\omega}_3){\omega}_a^{-1}
\ub{\tilde{\Phi}} \|^2 , \\ \label{eq:traj-dir-B}
B_a &= \alpha \langle \ub{\tilde{\Phi}} |
|\omega_2+\omega_3|^{-\frac{1}{2}} (-i\partial_{\omega_a})
|\omega_2+\omega_3|^{\frac{1}{2}} \ub{\tilde{\Phi}} \rangle .
\end{align}\end{subequations}

Evaluating them for the Gaussian form
\eqref{eq:wdw-gauss} of $\ub{\tilde{\Phi}}$ and taking
the limit $\sigma_a\to 0$ (for both $a=2,3$), we
obtain the following trajectory
\begin{equation}\label{eq:gauss-traj}
\langle \ub{\tilde{\Phi}} | \ln(\hat{v}_a)_{v_1}
\ub{\tilde{\Phi}} \rangle =
\bigg[\frac{\omega_1(\omega_2^\star,\omega_3^\star)}
{\omega_a^\star}\bigg]^2 \ln v_1 + \alpha\beta^a
\end{equation}
which agrees with the classical one. This result
implies that in the WDW theory the singularities of
the vacuum Bianchi I universe are not resolved
dynamically (in the sense of the trajectories defined
by the expectation values on semiclassical states).
Actually, the lack of singularity resolution is a
general property of all the states for which the
coefficients $A_a$ and $B_a$ defined in Eq.
\eqref{eq:traj-dir} are finite.

To analyze the behavior of the dispersions
$\langle\Delta\ln(\hat{v}_a)_{v_1}\rangle$, we first
find the expectation values of
$\ln^2(\hat{v}_a)_{v_1}$ in a way similar to the
derivation of Eq. \eqref{eq:wdw-traj-shape}. They read
\begin{equation}\label{eq:wdw-sqr-traj}
\langle \ub{\tilde{\Phi}} | \ln^2(\hat{v}_a)_{v_1}
\ub{\tilde{\Phi}} \rangle = W_a \ln^2 v_1 + Y_a \ln
v_1+X_a ,
\end{equation}
where
\begin{subequations}\label{eq:sqr-dir}\begin{align}
\label{eq:traj-dir-W} W_a &= \|
\omega_1^2({\omega}_2,{\omega}_3){\omega}_a^{-2}
\ub{\tilde{\Phi}} \|^2 , \\\label{eq:traj-dir-Y} Y_a
&= -2 i \alpha \langle \ub{\tilde{\Phi}} |
|\omega_2+\omega_3|^{-\frac{1}{2}}
\omega_1({\omega}_2,{\omega}_3)
{\omega}_a^{-1} (\partial_{\omega_a})\nonumber\\
& \times
|\omega_2+\omega_3|^{\frac{1}{2}}
\omega_1({\omega}_2,{\omega}_3)
{\omega}_a^{-1} \ub{\tilde{\Phi}} \rangle
\\\label{eq:traj-dir-X}
X_a &= - \alpha^2\langle \ub{\tilde{\Phi}} |
|\omega_2+\omega_3|^{-\frac{1}{2}} \partial_{\omega_a}^2
|\omega_2+\omega_3|^{\frac{1}{2}} \ub{\tilde{\Phi}} \rangle .
\end{align}\end{subequations}
In Eq. \eqref{eq:traj-dir-Y}, there is no summation
over the indices $a$. Using the standard relation
$\langle\Delta\ln(\hat{v}_a)_{v_1}\rangle^2 =
\langle\ln(\hat{v}_a)_{v_1}^2\rangle -
\langle\ln(\hat{v}_a)_{v_1}\rangle^2$ we can easily
find the dispersions. In particular, we see
immediately that once these dispersions are finite in
some epoch, they remain so throughout all the
evolution. Furthermore, for states for which the
expectation values $B_a$ and $X_a$ [defined in Eqs.
\eqref{eq:traj-dir} and \eqref{eq:sqr-dir}] are
finite, the relative dispersions approach constant
values in the large $v_1$ limit, values which are
determined by the relative dispersions of
$\omega_1^2({\omega}_2,{\omega}_3){\omega}_a^{-2}$:
\begin{equation}\label{eq:wdw-rel-disp}
\lim_{v_1\to\infty} \frac{\langle
\Delta\ln(\hat{v}_a)_{v_1} \rangle}{\langle
\ln(\hat{v}_a)_{v_1} \rangle} = \frac{\langle \Delta
[\omega_1^2({\omega}_2,{\omega}_3) {\omega}_a^{-2}]
\rangle} {\langle
\omega_1^2({\omega}_2,{\omega}_3){\omega}_a^{-2}
\rangle} .
\end{equation}
Here we have used the shorthand $\langle\hat{O}\rangle
:= \langle \ub{\tilde{\Phi}} | \hat{O}\,
\ub{\tilde{\Phi}} \rangle$ for any operator $\hat{O}$.

\subsection{$b$-representation}\label{sec:WDW-brep}

In the loop quantization, the evolution with respect
to the internal time $v_1$, analog to the one
constructed above for the WDW theory, fails to be
unitary. Nevertheless, the use of the momentum
conjugate to $v_1$ as internal time provides a good
notion of unitary evolution. Let us study this choice
of time in the WDW quantization as well, both for
completeness and in order to introduce the procedure
in a simple setting, where the difficulties inherent
to the more complicated nature of the LQC model are
absent.

Given the classical variable $v_i$ introduced in Eq.
\eqref{eq:v-def}, one can define the conjugate
momentum
\begin{equation}\label{eq:b-def}
b_i:=\sqrt{\Delta} c^i |p_i|^{-\frac{1}{2}},
\end{equation}
such that
\begin{equation}\label{eq:bv-poiss}
\{b_i,v_j\} = 2\delta_{ij}.
\end{equation}

In the WDW quantum theory, the unitary transformation
between the ``position'' and ``momentum''
representations is given by the Fourier transform:
\begin{equation}\label{eq:v-b-trans}
[{\ub{\mathcal{F}}}\psi](b_i) =
\frac1{2\sqrt{\pi}}\int_{\re} \rd v_i \,\psi(v_i)
e^{-\frac{i}{2}v_ib_i}.
\end{equation}
It is worth emphasizing that this is a unitary
transformation from the Hilbert space of square
integrable functions in the $v_i$-representation to
the \emph{same Hilbert space} in the
$b_i$-representation:
\begin{equation}
\ub{\mathcal{F}}: \ub{\Hil}{}_{\kin}^i=
L^2(\re,\rd v_i)\to\ub{\tilde\Hil}{}_{\kin}^i=
L^2(\re,\rd b_i)
\end{equation}

Under this transformation the elementary kinematical
operators transform as
\begin{subequations}\label{eq:ops-b}\begin{align}
\hat{v}_i\ &\to\ 2i\partial_{b_i},  &
\partial_{v_i}\ &\to\ i\hat{b}_i/2,
\tag{\ref{eq:ops-b}}\end{align}\end{subequations}
where $\hat{b}_i$ acts in the new representation as a
multiplication operator. Hence, the transformed of
$\widehat{\ub\Theta}_i$ is
\begin{equation}\label{eq:Th-trans}
\ub{\mathcal{F}}(\widehat{\ub{\Theta}}_i)=
-i\sqrt{3}\Delta (1+2b_i\partial_{b_i}),
\end{equation}
which coincides with the original operator
$\widehat{\ub\Theta}_i$ (up to a sign), both being
defined in identical Hilbert spaces. Therefore, in the
WDW quantum theory, working in the
$b_i$-representation is completely equivalent to
working in the $v_i$-representation. In particular, we
can regard $b_1$ as the internal time, change the
representation only in the direction 1, and define the
Hamiltonian constraint in the kinematical Hilbert
space $\ub{\tilde\Hil}{}_{\kin}^{1,+}\otimes_a
\ub{\Hil}^{a,+}_{\kin}$ ($a=2,3$), where
$\ub{\tilde\Hil}{}_{\kin}^{1,+}=L^2(\re^+,\rd b_i)$
\cite{Fn}, by replacing the operator
$\widehat{\underline\Theta}_1$ with the operator
$\ub{\mathcal{F}}(\widehat{\underline{\Theta}}_1)$ in
Eq. \eqref{CWDW}. Hence, we can repeat exactly the
construction introduced in Subsec.
\ref{sec:WDWphys-evo} substituting $v_1$ by $b_1$ and
$\ub{e}_{\omega_1}(v_1)$ by $\ub{e}_{-\omega_1}(b_1)$.

Whereas the $b_i$-representation does not introduce
any novelty or advantage in the WDW quantization in
comparison with the $v_i$-representation, we will see
in Sec. \ref{sec:lqc-b-rep} that there is a big
difference between both approaches in the LQC
quantization.

\section{Wheeler-DeWitt limit of the loop states}
\label{sec:wdw-limit}

The comparison of Eqs. \eqref{physpaWDW} and
\eqref{phystawdw} with Eqs.
\eqref{eq:fdec}-\eqref{physpa} shows that the physical
Hilbert spaces of the LQC and WDW quantizations, as
well as the structure of the corresponding wave
functions, are identical. The difference between both
quantizations is captured in the different form that
the eigenfunctions of the operators
$\widehat{\Theta}_i$ and $\widehat{\ub{\Theta}}_i$
possess. On the other hand (as it will be shown in
Subsec. \ref{sec:WDWlim-eigenf}), the LQC
eigenfunctions converge for large $v_i$ to some
combinations of their WDW analogs. This feature allows
one to regard the dynamics of the LQC universe as
certain form of ``scattering'' of WDW states, incoming
from a distant past and outgoing to a distant future
\cite{kp-disp}. Mathematically, this behavior is
described by the analog of a scattering matrix
$\hat{\rho}_{s}$ acting on the incoming WDW state:
\begin{subequations}\begin{align}\label{eq:scatt-def}
|\ub{\Phi}\rangle_{\rm out} &= \hat{\rho}_{s}
|\ub{\Phi}\rangle_{\rm in},\\
(\ub{e}_{\omega_2},\ub{e}_{\omega_3}|\hat{\rho}_{s}|
\ub{e}_{\omega'_2},\ub{e}_{\omega'_3})
&= \hat{\rho}_{1}(\omega_1(\omega_2,\omega_3),
\omega_1(\omega'_2,\omega'_3)) \notag \\
&\times \hat{\rho}_{2}(\omega_2,\omega'_2)
\hat{\rho}_{3}(\omega_3,\omega'_3)
\\
\hat{\rho}_{i}(\omega_i,\omega'_i) &:=
(\ub{e}_{\omega_i}|\hat{\rho}_{i}
|\ub{e}_{\omega'_i}) .
\end{align}\end{subequations}
In turn, the matrices $\hat{\rho}_{i}$ are
determined by the WDW limit of the eigenfunctions of
$\widehat{\Theta}_i$. The exact form of this
limit will be investigated in Subsec.
\ref{sec:WDWlim-eigenf}. The result will be applied in
Subsec. \ref{sec:wdw-limit-phys} to describe the limit
of physical states. Finally, Subsec.
\ref{sec:wdw-limit-disp} deals with the effect of the
commented scattering on the dispersion of
observables.

\subsection{Limit of the eigenfunctions}
\label{sec:WDWlim-eigenf}

Let us restrict ourselves to one superselection sector
(defined in Subsec. \ref{sec:fr-Hph}), e.g. that
corresponding to functions supported on the lattice
$\lat^+_{\varepsilon_i}$. The eigenfunctions
$e^{\varepsilon_i}_{\omega_i}$ of $\widehat{\Theta}_i$
with eigenvalue $\omega_i$ were provided already in
Ref. \cite{mmp} [see Eq. (45)], although in the form
presented there one cannot easily determine their
large $v_i$ behavior. In order to find it, we have to
analyze the operator $\widehat{\Theta}_i$ itself. Its
properties were discussed in detail in Ref. \cite{mmp}
as part of the proof of its self-adjointness. First,
all of its eigenspaces are one-dimensional, and the
complete solution is determined by an initial value
$e^{\varepsilon_i}_{\omega_i}(\varepsilon_i)$. In
particular, the freedom in the choice of a global
phase for the eigenfunctions is removed by demanding
the positivity of this initial value
$e^{\varepsilon_i}_{\omega_i}(\varepsilon_i)$. Second,
as discussed in Subsec. \ref{sec:fr-Hph}, one can
split the support into two subsemilattices
${}^{(4)}\!\lat_{\tilde{\varepsilon}_i}^+$, with
$\tilde{\varepsilon}_i\in\{\varepsilon_i,
\varepsilon_i+2\}$. The restriction of
$e^{\varepsilon_i}_{\omega_i}$ to each subsemilattice
is an eigenfunction of the operator
$\widehat{\Theta}_i^2$ with eigenvalue $\omega_i^2$:
\begin{equation}\label{eq:eig-eq-restr}
\widehat{\Theta}_i^2
e^{\tilde{\varepsilon}_i}_{\omega_i} = \omega_i^2
e^{\tilde{\varepsilon}_i}_{\omega_i}, \quad
e^{\tilde{\varepsilon}_i}_{\omega_i} :=
e^{\varepsilon_i}_{\omega_i}|_{{}^{(4)}\!
\lat_{\tilde{\varepsilon}_i}^{+}}.
\end{equation}
The operator $\widehat{\Theta}_i^2$, in turn, is a
second order difference operator with real
coefficients (similar in structure to the evolution
operator defined in Ref. \cite{aps-imp}). It also has
the property that all its eigenspaces are
one-dimensional and the eigenfunctions are determined
just by their initial value (at
$v_i=\tilde{\varepsilon}_i$).

To check the existence of the WDW limit of the
eigenfunctions, we implement the method used in Ref.
\cite{kp-posL}:
\begin{enumerate}[(i)]
\item First, we represent the values of the
eigenfunction at two consecutive points of
${}^{(4)}\!\lat_{\tilde{\varepsilon}_i}^+$ by vectors
$\vec{\psi}(v)$ of coefficients of its decomposition
in terms of the WDW eigenfunctions
$\ub{e}_{\pm|\omega_i|}$ evaluated at this pair of
points.
\item Next, we rewrite the eigenfunction equation as the
first order one acting on the vectors $\vec{\psi}(v)$.
It has a form of a $2\times 2$ matrix [denoted from
now on as $\boldsymbol{B}(v)$].
\item Finally, we calculate the asymptotic expansion of
the matrix $\boldsymbol{B}$
in the large $v$ limit.
\end{enumerate}
An explicit calculation shows that the considered
matrix is of the form $\boldsymbol{B}(v) = \id +
\boldsymbol{O}(v^{-2})$. Thus, there exists a well
defined limit
\begin{equation}
\vec{\psi} := \lim_{v\to\infty} \vec{\psi}(v) .
\end{equation}
This immediately implies the convergence to a
combination of functions in the WDW basis, represented
by the coefficient vector $\vec{\psi}$. The form of
$\boldsymbol{B}$ implies also that the rate of
convergence of $\vec{\psi}(v)$ is at least of order
$1/v$.

Actually, the fact that the operator
$\widehat{\Theta}_i^2$, after a suitable change of
representation, differs from the isotropic evolution
operator of Ref. \cite{aps-imp} just by a compact term
(see Ref. \cite{mmp}) allows us to apply here the
numerical results of that reference. They show that
the convergence is even faster, namely
\begin{equation}\label{eq:conv-rate}
\vec{\psi}(v) = \vec{\psi} +
\boldsymbol{O}(v^{-\frac{5}{2}}).
\end{equation}
Furthermore, from the reality of
$\widehat{\Theta}_i^2$, it follows that the incoming
and outgoing WDW plane waves contribute equally to the
limit, that is
\begin{equation}\label{eq:lqc-wdw-conv}
\vec{\psi} = r\left[ \begin{array}{c} e^{i\phi(\omega_i)} \\
e^{-i\phi(\omega_i)}\end{array} \right].
\end{equation}

To determine the normalization factor $r$ we note that
$\lim_{v\to\infty}(\partial_{v}\ub{e}_{\omega_i}(v))
/\ub{e}_{\omega_i} (v)=0$. This fact, together with
the sufficiently fast rate of convergence
\eqref{eq:conv-rate}, imply that the
(kinematical) norms of both
$e^{\tilde{\varepsilon}_i}_{\omega_i}$ and its WDW
limit (denoted here as $\ub{e}^{\tilde{\varepsilon}_i}_{\omega_i}$) satisfy
the relation $8\| e^{\tilde{\varepsilon}_i}_{\omega_i}\|^2 
= \| \ub{e}^{\tilde{\varepsilon}_i}_{\omega_i}\|^2$. As a
consequence
\begin{equation}\label{eq:r}
  r(\omega_i)=\sqrt{2}z_i ,
\end{equation}
where $z_i$ is a global phase which equals $1$ for
$\tilde{\varepsilon}_i\leq2$ and $-i\sgn(\omega_i)$
otherwise.

The phase shift $\phi(\omega_i)$ has a nontrivial
dependence on $\omega_i$ and needs to be found
numerically. Luckily, the similarity of the operator
$\widehat{\Theta}_i^2$ with the evolution operator of
an isotropic universe allows us again to apply
directly the methods of Ref. \cite{kp-disp}. The
result is the following:
\begin{equation}\label{eq:alpha-form}
\phi(\omega_i) = (\ln|\omega_i| + a)(|\omega_i| + b)
+ c_{\tilde{\varepsilon}_i} +
R_{\tilde{\varepsilon}_i}(|\omega_i|)  ,
\end{equation}
where $a$, $b$ and $c_{\tilde{\varepsilon}_i}$ are
constants, $\lim_{\omega_i\to\infty}
R_{\tilde{\varepsilon}_i}(|\omega_i|) = 0$, and the
dependence on $\tilde{\varepsilon}_i$ enters only in
the constant term and in the remnant part. Hence, for
large $|\omega_i|$ the terms that affect the position
and dispersion of the wave packet do not depend on the
value of $\tilde{\varepsilon}_i$.

\begin{figure}[htb!]
\includegraphics[width=3.2in]{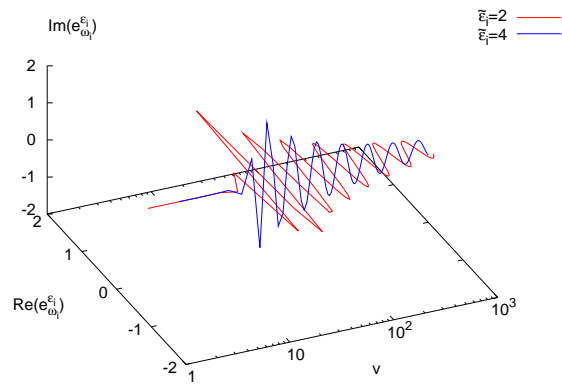}
\caption{An example of eigenfunction of the operator
$\widehat{\Theta}_i$, corresponding to the eigenvalue
$\omega_i=100$ and the superselection sector
$\varepsilon_i=2$. The blue line (located on the imaginary plane) 
shows the part supported on the subsemilattice
${}^{(4)}\!\lat_{\tilde{\varepsilon}_i=4}^+$, whereas
the red line (real plane) is the part supported on
${}^{(4)}\!\lat_{\tilde{\varepsilon}_i=2}^+$.}
\label{fig:eigenf}
\end{figure}

At this point, we can already write the exact form of
the scattering matrix $\hat{\rho}_i$ defined for just
one subsemilattice. It reads:
\begin{equation}
\hat{\rho}_i(\omega_i,\omega'_i) = e^{-2i\phi(\omega_i)}
\delta(\omega_i+\omega'_i) .
\end{equation}
Let us remember that the parts supported on different
subsemilattices (individually of constant phase) are
shifted in phase by $\pi/2$. As a consequence, the
common WDW limit for both of them does not exist (see
Fig.~\ref{fig:eigenf}). Therefore, one cannot write an
explicit form for the scattering matrix $\hat{\rho}_i$
on the entire lattice $\lat^+_{\varepsilon_i}$.
Nonetheless, the differences between subsemilattices
manifest themselves only through the constant phase
shift and the remnant decaying for large $\omega_i$.
As a consequence, when we consider the properties of
asymptotic wave packets, peaked around large
$\omega_a$, we can safely restrict the studies just
to one subsemilattice.

\subsection{Physical states}\label{sec:wdw-limit-phys}

The WDW limit of the eigenfunctions found in the
previous subsection can be now applied in the analysis
of the physical states. We start with a general state
$\tilde{\Phi}$, again restricting the study just to
particular subsemilattices
${}^{(4)}\!\lat_{\tilde{\varepsilon}_i}^+$. In order
to find the limit of the corresponding wave function,
we simply replace the basis functions
$e^{\tilde{\varepsilon}_i}_{\omega_i}$ in Eq.
\eqref{eq:fdec} with their large $v_i$ limits
\begin{equation}
e^{\tilde{\varepsilon}_i}_{\omega_i} \to r [
e^{i\phi(\omega_i)}\, \ub{e}_{\omega_i} +
e^{-i\phi(\omega_i)}\, \ub{e}_{-\omega_i} ] .
\end{equation}
Upon this replacement, the wave function
$\Phi(\vec{v})$ is transformed into the function
\begin{equation}\label{eq:lim-split}\begin{split}
\ub{\Phi}(\vec{v}) = \sum_{s_2,s_3=\pm 1} \,\, &
\int_{\re^2} \rd\omega_2\rd\omega_3\,
\tilde{\Phi}_{\vec{s}}(\omega_2,\omega_3) \\
&\times
\ub{e}_{\omega_{\vec{s}}(\omega_2,\omega_3)}(v_1) \,
\ub{e}_{\omega_2}(v_2) \, \ub{e}_{\omega_3}(v_3) ,
\end{split}\end{equation}
where $\vec{s}:=(s_2,s_3)$,
$\omega_{\vec{s}}(\omega_2,\omega_3) :=
\omega_1(s_2\omega_2,s_3\omega_3)$, and
\begin{eqnarray}
&&\tilde{\Phi}_{\vec{s}}(\omega_2,\omega_3) =
2\sqrt{2}\sum_{s_1=\pm 1}\tilde{\Phi}
(s_1s_2\omega_2,s_1s_3\omega_3)
e^{is_1\phi(\omega_{s_1\vec{s}})} \nonumber\\
&&\times s_1^{\frac{3-z_1^2-z_2^2-z_3^2}{2}}z_1
\prod_{a=2}^3 s_a^{\frac{1-z_a^2}{2}}
z_a e^{is_{1}s_a\phi(s_1s_a\omega_a)} ,
\end{eqnarray}
where $z_i^2$ equals $1$ for
$\tilde{\varepsilon}_i\leq2$ and $-1$ otherwise. Note
that, in order to obtain the above expression, we have
taken into account that $\omega_1(-\omega_2,-\omega_3)
= -\omega_1(\omega_2,\omega_3)$ and
$r(s_j\omega_i)=s_j^{(1-z_i^2)/2}\sqrt{2}z_i$
(with $\omega_1=\omega_{\vec{s}}$).

If we compare Eq. \eqref{eq:lim-split} with Eq.
\eqref{phystawdw} we see that each term, corresponding
to a particular set of values of $s_a$, has a form
very similar to that of a WDW state with spectral
profile $\ub{\tilde{\Phi}} = \tilde{\Phi}_{\vec{s}}$,
the only difference being the replacement of the
function $\omega_1$ with $\omega_{\vec{s}}$ in the
index of the basis functions. As a consequence, each
of those terms can be (independently) considered as a
state defined within a certain analog of the WDW
quantum theory described in Sec.~\ref{sec:wdw-analog}.
Such an analog inherits all the properties and
structure of the original theory, except for the
transformation of $\omega_1\to\omega_{\vec{s}}$
pointed out above. In particular, the inner product
and the definitions of all the observables remain
unmodified.

This correspondence allows us to apply directly the
definitions \eqref{obs} of the observables
$\ln(\hat{v}_a)_{v_1}$ to the terms
$\tilde{\Phi}_{\vec{s}}$. Their expectation values
(calculated on each term independently) satisfy an
analog of the equation \eqref{eq:wdw-traj-shape} with
the direction coefficient $A_a$ replaced with
\begin{equation}
A_{a,\vec{s}}:=s_a\|
\omega_{\vec{s}}({\omega}_2,{\omega}_3){\omega}_a^{-1}
\ub{\tilde{\Phi}} \|^2 .
\end{equation}
To select the terms $\tilde{\Phi}_{\vec{s}}$ which
actually contribute to the investigated limit we note
that, if the state is localized (in the sense that it
remains peaked around the trajectories defined by the
expectation values of the considered observables),
only those terms corresponding to an $\vec{s}$ for
which $A_{a,\vec{s}}$ is strictly positive will have a
significant contribution in the regime when the three
$v_i$'s are all large. This requirement is satisfied
only by the term with $s_2=s_3=1$.

The surviving term encodes the state of the genuine
(untransformed) WDW theory. Therefore, the large $v_i$
limit of a localized state $\tilde{\Phi}$ is simply
given by a WDW state of spectral profile
\begin{eqnarray}\label{eq:wdw-limit}
\ub{\tilde{\Phi}}(\omega_2,\omega_3) &=&
2\sqrt{2}\sum_{s=\pm 1}
s^{\frac{3-z_1^2-z_2^2-z_3^2}{2}}
\tilde{\Phi}(s\omega_2,s\omega_3) \nonumber \\
&&\times \prod_{i=1}^3 z_i \,
e^{is\phi(s\omega_i)},
\end{eqnarray}
where $\omega_1$ is again related to $\omega_2$ and
$\omega_3$ via Eq. \eqref{om-func}.

Let us recall at this stage that the variable $v_1$
plays the role of time; therefore it is proper to
introduce a decomposition of the state into positive
and negative frequency parts, corresponding to
$\omega_1>0$ and $\omega_1<0$ respectively.
Physically, these components can be interpreted as
moving forward and backward in time. An analog
correspondence can be applied to other directions,
defining the splitting into expanding and contracting
components. Then, the change of sign in $\omega_a$
corresponds to a parity reflection in $x_a=\ln(v_a)$.
As a consequence, the transformation $\{\omega_i\}\to
\{-\omega_i\}$, which we have recognized as a symmetry
of Eq. \eqref{om-func} in our previous discussion, is
the analog of the full PT (parity/time inversion)
transformation.

Let us now consider a LQC state with profile
$\tilde{\Phi}(\omega_2,\omega_3)$ which is localized
in the sense explained above, and restrict our
considerations to the corresponding WDW state with
profile $\ub{\tilde{\Phi}}(\omega_2,\omega_3)$. For
convenience, in the following we will call
$2\sqrt{2}\tilde{\Phi}(\omega_2,\omega_3)$ \emph{the
reference state}. The expectation values of
$\ln(\hat{v}_a)_{v_1}$ follow then the trajectory
\eqref{eq:wdw-traj-shape} for some constants $A_a$ and
$B_a$. Furthermore, from Eq. \eqref{eq:wdw-limit}, one
can immediately see that the WDW state consists of two
parts: one with the same parity/time orientation as
the reference state ($s=1$) and another which is PT
reflected ($s=-1$). We denote them, respectively, by
$\ub{\tilde{\Phi}}_+$ and $\ub{\tilde{\Phi}}_-$. If
the reference state has a definite time orientation
(only one sign of $\omega_1$ contributes) then the
distinguished components are, respectively, the part
comoving in time with the reference state and the
time-reflected part. Since the norms of both of these
components are equal, any wave packet moving backward
in time is fully reflected into a packet moving
forward. This shows the presence of a bounce also in
the internal time direction.

In order to compare the trajectories of the
expectation values of $\ln(\hat{v}_a)$ in the
considered components with those of the reference
state, we note that, since their spectral profiles are
related just by a rotation, and by the reflection in
$\omega_a$ together with a possible change of sign in
the case of $\ub{\tilde{\Phi}}_-$, the directional
coefficients $A_a$ are the same for these component
states and the reference state. The only change is in
the parameter $B_a$. As a consequence, the
trajectories are just shifted with respect to the
trajectory of the reference state.

In the particular case of Gaussian reference states
which are sharply peaked around large
$\omega^\star_a$,
\begin{equation}\label{eq:gauss-state}
\tilde{\Phi}(\omega_2,\omega_3) =
\frac{K}{\sqrt{|\omega_2+\omega_3|}}\prod_{a=2}^{3}
e^{-\frac{(\omega_a-\omega_a^\star)^2}{2\sigma_a^2}}
e^{i\beta^a\omega_a}
\end{equation}
(where $K$ is chosen so that $\ub{\tilde{\Phi}}$ has
unit norm), we can provide a quantitative estimation
of the trajectory shift. Namely, for such states the
phase rotation $\phi(\omega_i)$ is well approximated
by its first order expansion (with coefficients $D$
and $E$ which depend on $\omega^\star_i$)
\begin{equation}\begin{split}
\phi(\omega_i) &\approx
D(\omega^\star_i)(\omega_i-\omega^\star_i)
+E(\omega_i^\star),
\\
D(\omega^\star_i) &=
\sgn{(\omega^\star_i)}(1+a+\ln|\omega^\star_i|),
\end{split}\end{equation}
where $a$ is the constant given in Eq.
\eqref{eq:alpha-form}. In view of this approximation,
one can almost directly repeat the calculations of
Subsec. \ref{sec:WDWphys-evo} to find the formula
equivalent to Eq. \eqref{eq:wdw-traj-shape}. The
resulting trajectories are
\begin{equation}\label{eq:limit-trajs}\begin{split}
[\ln v_a](v_1) &=
\bigg[\frac{\omega_1(\omega_2^\star,\omega_3^\star)}
{\omega^\star_a}\bigg]^2
[\ln v_1 - \alpha D(\omega^\star_1)] \\
&+ \alpha[D(\omega^\star_a)\pm\beta^a].
\end{split}\end{equation}
Generically, these trajectories are disjoint, both
between them and with respect to the trajectory of the
reference state. Nonetheless, note that the two
considered trajectories coincide in the case
$\beta^a=0$.

\subsection{The dispersions}\label{sec:wdw-limit-disp}

In the discussion above, we have taken the limit
$\sigma_a\to 0$ ($a=2,3$) and considered the linear
approximation to the variation of the rotation phase
$\phi(\omega_i)$. This essentially removes all the
information about the behavior of the dispersions.
However, in the scenario described in the previous
subsection, where the LQC dynamics can be (in an
asymptotic sense) viewed as the reflection of the WDW
wave packet $\ub{\tilde\Phi}_-$, which moves backward
in time and is contracting, into the moving forward
and expanding wave packet $\ub{\tilde\Phi}_+$, it is
important to ask how much the dispersion of
$\ub{\tilde\Phi}_+$ grows in comparison to the one of
$\ub{\tilde\Phi}_-$ (or vice versa). For the isotropic
model with massless scalar field, restrictive bounds
on possible dispersion-growth have been found. This
result can be extended by employing the exact triangle
inequalities \cite{kp-disp} which involve the
dispersion in $\ln|\omega_a|$ and $\ln(\hat{v}_a)$
(for brevity, we suppress the subindex $v_1$ in the
latter of these operators).

In the model considered here, given that the analysis
of the WDW analog shows that the \emph{relative}
dispersions in $\ln(\hat{v}_a)$ are the ones which
approach constant (nonzero) values for large $v_1$
[see Eq. \eqref{eq:wdw-rel-disp}], we are interested
in finding a weaker relation, involving exactly these
relative quantities.

We begin by recalling that the complete wave function
$\Phi$ is supported on the product of three
semilattices $\lat_{\varepsilon_i}^+$, each of which
is in turn the union of two semilattices of step four,
${}^{(4)}\lat_{\tilde{\varepsilon}_i}$ (where
$\tilde{\varepsilon_i} \in
\{\varepsilon_i,\varepsilon_i+2\}$). We then divide
the support into eight sectors, corresponding to
the eight products of the chosen semilattices of step
four, and consider the respective restrictions of
$\Phi$ to each of them. In each sector, we apply the
scattering scheme defined at the beginning of this
section. In that scheme, the wave packet
$\ub{\tilde\Phi}_-$ is transformed into
$\ub{\tilde\Phi}_+$ via the unitary rotation
\begin{equation}\label{eq:rotation}
  U = \prod_{i=1}^3 e^{2i\phi(\omega_i)},
\end{equation}
the reflection in the signs of $\omega_a$, and a
possible change of global sign.

It seems reasonable to restrict our considerations to
states such that the \emph{each} of the components
$\ub{\tilde\Phi}_-$ corresponding to the sectors
introduced above is such that the coefficients $A_a$,
$B_a$, $W_a$, $Y_a$, and $X_a$, defined in
Eqs.~\eqref{eq:traj-dir} and \eqref{eq:sqr-dir}, are
finite. This ensures that each of these components has
a well defined associated trajectory
\eqref{eq:wdw-traj-shape} and a finite dispersion for
every finite $v_1$. In addition, Eq.
\eqref{eq:wdw-rel-disp} is satisfied.

Neither the rotation \eqref{eq:rotation}, nor the
reflection in the signs of $\omega_a$, nor the
possible change in sign of the wave function will
change the expectation value or the dispersion of the
operator $\omega_1^2(\omega_2,\omega_3)\omega_a^{-2}$.
Therefore, if the coefficients \eqref{eq:traj-dir} and
\eqref{eq:sqr-dir} are finite also for
$\ub{\tilde\Phi}_+$, Eq.~\eqref{eq:wdw-rel-disp}
implies immediately that the following holds
\begin{equation}\label{eq:lim-disp-eq}
\frac{\langle\Delta\ln
(\hat{v})_a\rangle_+}{\langle\ln (\hat{v})_a\rangle_+}
= \frac{\langle\Delta\ln
(\hat{v})_a\rangle_-}{\langle\ln (\hat{v})_a\rangle_-}
,
\end{equation}
where $\langle\,\cdot\,\rangle_{\pm}$ denote
expectation values on $\ub{\tilde\Phi}_{\pm}$,
respectively.

To prove that the finiteness of $A_a$, $B_a$,
$W_a$, $Y_a$, and $X_a$ for $\ub{\tilde\Phi}_-$
implies the finiteness of these coefficients for
$\ub{\tilde\Phi}_+$, we apply similar methods to those
proposed by Kami\'nski and Pawlowski \cite{kp-disp}.
Since $A_a$ and $W_a$ are equal for both components,
the only ones that require detailed analysis are
$B_a$, $Y_a$, and $X_a$. We first recall that
\begin{subequations}\label{eq:coeff-reorg}\begin{align}
&B_a = \langle \mathcal{D}_a \rangle, \qquad Y_a =
\langle \mathcal{D}'_a \rangle \\
&X_a-B_a^2 = \langle \Delta \mathcal{D}_a \rangle^2 =:
\sigma_{\mathcal{D}_a}^2,
\end{align}\end{subequations}
where
\begin{subequations}\label{eq:difXB}\begin{align}
\mathcal{D}_a &=
-i\alpha|\omega_2+\omega_3|^{-\frac{1}{2}}
(\partial_{\omega_a})|\omega_2+\omega_3|^{\frac{1}{2}} ,
\\ \mathcal{D}'_a &=
2\omega_1(\omega_2,\omega_3)\omega_a^{-1} \,
\mathcal{D}_a \,
\omega_1(\omega_2,\omega_3)\omega_a^{-1}.
\end{align}\end{subequations}
Knowing the relation between $\ub{\tilde\Phi}_-$ and
$\ub{\tilde\Phi}_+$, we can find the relations between
their corresponding expectation values
\eqref{eq:coeff-reorg}. They are
\begin{subequations}\label{eq:exp-disp-trans}\begin{align}
\langle \mathcal{D}_a \rangle_+ &= - \langle
\mathcal{D}_a \rangle_- -2\alpha\sum_{i=1}^3 \langle
[\partial_{\omega_a}
\phi(-\omega_i)] \rangle_- \label{eq:rel-exp}\\
\langle \mathcal{D}'_a \rangle_+ &= - \langle
\mathcal{D}'_a \rangle_- -4\alpha\sum_{i=1}^3 \langle
\frac{\omega_1^2}{\omega_a^{2}} [\partial_{\omega_a}
\phi(-\omega_i)] \rangle_-
\label{eq:rel-exp2}\\
\sigma_{\mathcal{D}_a+} &\leq \sigma_{\mathcal{D}_a-}
+2\alpha\sum_{i=1}^3
\langle\Delta[\partial_{\omega_a}\phi(-\omega_i)]\rangle_-
\label{eq:rel-disp}
\end{align}\end{subequations}
Actually, it is possible to estimate the terms related
with $\partial_{\omega_a}\phi(-\omega_i)$ employing
that the function $\phi(\omega)$ possesses the
following properties \cite{kp-disp}:
\begin{subequations}\begin{align}
&|\partial_{\omega}\phi (\omega)| \leq C_1
\left|\ln|\omega|\right| + C_0 , \label{eq:phi-1}\\
&|\omega\partial_{\omega}^2\phi (\omega)|
\leq C_2 \label{eq:phi-2} ,
\end{align}\end{subequations}
where $C_0$, $C_1$, and $C_2$ are (positive)
finite constants, which however may depend on the
value of the subsemilattice label
$\tilde{\varepsilon}_i$ and, in particular, may not
have a global bound (in the whole interval of
variation of this label).

These inequalities can be next used to relate the
terms in Eq. \eqref{eq:exp-disp-trans} with the
expectation values and dispersions of the operators
$\ln|\omega_i|$. In order to do so, let us first
define the multiplicative operators
\begin{subequations}\begin{align}
w_a &:= \omega_1(\omega_2,\omega_3)
\omega_a^{-1} , \\
\Omega^{(n)}_a &:= w_a^n
\ln|\omega_1(\omega_2,\omega_3)| , \\
\Sigma^{(n)}_a &:= w_a^n \ln|\omega_a| .
\end{align}\end{subequations}
In the case of relation \eqref{eq:rel-exp}, the last
term is bounded as follows
\begin{subequations}\label{eq:bound-exp}\begin{align}
& \left| \sum_{i=1}^3 \langle
[\partial_{\omega_a}\phi(-\omega_i)] \rangle_- \right|
\tag{\ref{eq:bound-exp}}\\
&\qquad \leq C_1 \left[ \langle | \ln{|\omega_a|}
|\rangle_- + \langle |\Omega^{(2)}_a| \rangle_-
\right] + C_0(1+\langle w_a^2 \rangle) , \notag
\end{align}\end{subequations}
whereas for the term in Eq. \eqref{eq:rel-exp2} one
has
\begin{subequations}\label{eq:bound-exp2}\begin{align}
&\left| \sum_{i=1}^3 \langle
\frac{\omega_1^2}{\omega_a^{2}}
[\partial_{\omega_a}\phi(-\omega_i)] \rangle_- \right|
\tag{\ref{eq:bound-exp2}}\\
&\qquad \leq C_1 \left[ \langle| \Sigma^{(2)}_a
|\rangle_- + \langle|\Omega^{(4)}_a|\rangle_- \right]
+ C_0\left[\langle w_a^2\rangle+\langle
w_a^4\rangle\right] , \notag
\end{align}\end{subequations}
Similarly, the sum in \eqref{eq:rel-disp} satisfies
\begin{equation}\label{eq:bound-disp}\begin{split}
&\sum_{i=1}^3 \langle\Delta[\partial_{\omega_a}
\phi(-\omega_i)]\rangle_- \\
&\qquad \leq C_2\left[
\langle\Delta\ln|\omega_a|\rangle_- + \langle\Delta
\Omega^{(2)}_a\rangle_- \right]
\end{split}\end{equation}

Suppose now that in the state $\ub{\tilde\Phi}_-$ the
dispersions and expectation values of $\ln|\omega_a|$,
$\Omega_a^{(2)}$, and $w_a^2$ are finite, as well as
the expectation values of $\Sigma^{(2)}_a$. This
immediately implies the finiteness of the right-hand
side of Eqs.
\eqref{eq:bound-exp}-\eqref{eq:bound-disp}
(including that of
$\langle|\Omega^{(4)}_a|\rangle_-$). From this and
the fact that $C_0,C_1,C_2<\infty$ it is
straightforward to check that, if the coefficients
$B_a$, $X_a$, and $Y_a$ corresponding to
$\ub{\tilde\Phi}_-$ are finite, so are the ones
corresponding to $\ub{\tilde\Phi}_+$. Therefore, we
conclude that Eq.~\eqref{eq:lim-disp-eq} is indeed
satisfied, as we wanted to prove.

It is worth noticing that, in the previous discussion,
the roles of $\ub{\tilde\Phi}_+$ and
$\ub{\tilde\Phi}_-$ can be interchanged, so that one
can instead impose mild conditions of the type
explained above on $\ub{\tilde\Phi}_+$ and ensure then
a good behavior for the relative dispersions
corresponding to $\ub{\tilde\Phi}_-$.

Finally, in Appendix \ref{app:unidis} we show that our
result \eqref{eq:lim-disp-eq} about the relative
dispersions can actually be extended to the case in
which one takes into consideration not just one
isolated sector, but the whole ensemble of the eight
sectors in which the LQC physical states admit a WDW
limit. In this way, we arrive at the following
conclusion. Consider a physical state described by the
wave function $\Phi$ supported on the product of
semilattices $\lat_{\varepsilon_1}^+ \times
\lat_{\varepsilon_2}^+ \times \lat_{\varepsilon_3}^+$.
Suppose that the restriction of $\Phi$ to the product
of subsemilattices
${}^{(4)}\lat_{\tilde{\varepsilon}_1}^+ \times
{}^{(4)}\lat_{\tilde{\varepsilon}_2}^+ \times
{}^{(4)}\lat_{\tilde{\varepsilon}_3}^+$ possesses a
WDW limit such that the component
$\ub{\tilde\Phi}_\pm$ moving forward/backward in time
has finite expectation values and dispersions for the
operators
\begin{equation}\label{eq:loc-set}
\ln(\hat{v}_a)_{v_1}, \ln|\omega_a|, \Omega^{(2)}_a ,
w_a^2.
\end{equation}
Suppose also that this state has finite expectation
values for the operators
\begin{equation}\label{eq:fin-set}
\Sigma^{(2)}_a.
\end{equation}
Then
\begin{enumerate}[(i)]
\item the corresponding component $\ub{\tilde\Phi}_\mp$
moving backward/forward in time has also finite
expectation values and dispersions with respect to the
operators \eqref{eq:loc-set}, as well as finite
expectation values for the operators
\eqref{eq:fin-set}, and
\item relation \eqref{eq:lim-disp-eq} holds for
the ensemble of the WDW limits corresponding to all
of the eight sectors defined by the restrictions to
the different subsemilattices, constructed to reflect the 
relevant features of a complete LQC state (see Appendix \ref{app:unidis} 
for the discussion).
\end{enumerate}

\section{Description on $v$-sections: Unitary evolution}
\label{sec:v-evo}

In the case of the WDW quantization, we introduced in
Subsec. \ref{sec:WDWphys-evo} the notion of evolution
by means of a family of observables which are related
via unitary transformations. In this section we will
analyze the possibility of performing an analogous
construction in the LQC model. First, in Subsec.
\ref{sec:vHil}, we will establish the relation between
the physical Hilbert space and an appropriate space of
``initial'' data defined on a single slice
$v_1=\const$. That relation will be used in Subsec.
\ref{sec:pre-v-obs} to construct a direct analog of
the family \eqref{obs}, which however fails to admit a
unitary relation. In Subsec. \ref{sec:v-obs}, certain
modification of the construction will allow us to
overcome this problem, although at the price of
loosing a neat physical interpretation of the selected
observables, which is recovered only in the large
$v_1$ limit. The modified observables are finally used
in Subsec. \ref{sec:v-num} to extract physical
predictions, which are presented in Subsec.
\ref{sec:v-results}.

\subsection{Time slices and associated Hilbert spaces}
\label{sec:vHil}

Let us start with the general form of the wave
function that represents the physical state, given by
Eq.~\eqref{eq:fdec}. In analogy with the procedure
explained in Subsec. \ref{sec:WDWphys-evo}, we choose
as the internal time the variable $v_1$ and define the
``initial data'' functions on each slice $v_1=\const$
in the following way
\begin{equation}\label{eq:phi-chi}
\Phi_{v_1}(v_2,v_3)\ =\ \int_{\re^2}
\rd\omega_2\rd\omega_3 \tilde{\Phi}_{v_1}
(\omega_2,\omega_3) e^{\varepsilon_2}_{\omega_2}(v_2)
e^{\varepsilon_3}_{\omega_3}(v_3) ,
\end{equation}
where the spectral profiles
$\tilde{\Phi}_{v_1}(\omega_2,\omega_3)$ of
$\Phi_{v_1}(v_2,v_3)$ belong to the slice Hilbert
spaces
\begin{equation}\label{eq:Hv-def}
\mathcal{H}_{v_1} := L^2(\re^2,
|\omega_2+\omega_3||e^{\varepsilon_1}_{\omega_1}(v_1)|^{-2}
\rd\omega_2\rd\omega_3),
\end{equation}
and are defined by the transformation
$P_{v_1}:\Hilpe\to\Hil_{v_1}$
\begin{align}\label{eq:chidef}
\tilde{\Phi}_{v_1}(\omega_2,\omega_3) &:=
P_{v_1}\tilde{\Phi}(\omega_2,\omega_3) \nonumber\\
&:= \tilde{\Phi}(\omega_2,\omega_3)
e^{\varepsilon_1}_{\omega_1}(v_1).
\end{align}

On any slice $v_1=\const$,
$e^{\varepsilon_1}_{\omega_1}(v_1)$ provides just a
function of $\omega_1$ which turns out to vanish in a
set of zero measure [see Eq. (45) in Ref. \cite{mmp}
for the details]. Therefore, the map $P_{v_1}$ is
unitary. This property, together with the fact that
both $e^{\varepsilon_2}_{\omega_2}(v_2)$ and
$e^{\varepsilon_3}_{\omega_3}(v_3)$ form bases of
their corresponding kinematical spaces $\mathcal
H_{\varepsilon_a}^+$, allows us to determine
$\tilde\Phi(\omega_2,\omega_3)$ from
$\Phi_{v_1}(v_2,v_3)$ (up to a zero measure set). As a
consequence, the projection on each $v_1$-slice
contains the same information as the entire physical
solution. However, one cannot write the inner product
of \eqref{eq:Hv-def} as an integral of
$\Phi_{v_1}(v_2,v_3)$ with well defined Lebesgue
measure. Owing to the dependence of
$|e^{\varepsilon_1}_{\omega_1}(v_1)|$ in $\omega_2$
and $\omega_3$, the inner product of
$\mathcal{H}_{v_1}$ is nonlocal when expressed in
terms of $v_2$ and $v_3$.

The unitary transformation $P_{v_1}$ allows us to
define a map between initial data spaces. Each state
on the physical Hilbert space $\Hilpe$ is associated,
through $P_{v_1}$, with a sequence of elements of the
slice spaces $\mathcal{H}_{v_1}$. Each sequence
consists in the chain of ``evolution steps''
enumerated by $v_1\in\lat_{\varepsilon_1}^+$. However,
the corresponding evolution is not unitary, because
under the identification of different slices, initial
data belonging to one of the spaces
$\mathcal{H}_{v_1}$ will in general not belong to the
others, since $|e^{\varepsilon_1}_{\omega_1}(v_1)|$
depends on $v_1$. In the next subsection we will try
to provide a more sophisticated notion of evolution
free of this problem by building a set of observables
analogous to the family \eqref{obs} that we
constructed for the WDW model.

\subsection{$v_1$-observables}
\label{sec:pre-v-obs}

Once we have introduced the Hilbert spaces
\eqref{eq:Hv-def}, and the transformations between
them and $\Hilpe$, we can follow the construction of
relational observables made in Subsec.
\ref{sec:WDWphys-evo}, starting from the kinematical
observables $\ln(\hat{v}_a)$ ($a=2,3$), which also act
as multiplication operators here. However, unlike the
WDW eigenfunctions $\ub{e}_{\omega_1}(v_1)$, the
eigenfunctions $e^{\varepsilon_1}_{\omega_1}(v_1)$
have a phase which is $v_1$-independent  separately on
each of the subsemilattices
${}^{(4)}\!\lat_{\varepsilon_1}^+$ and
${}^{(4)}\!\lat_{\varepsilon_1+2}^+$, with a global
phase shift of $\pi/2$ between them (see Subsec.
\ref{sec:fr-Hph}). Therefore the transformation
between $\Hil_{v_1}$ and $\Hil'$ analogous to
\eqref{eq:wdw-h'tr} essentially removes all the
information from the state. As a consequence, the
observables constructed in this way do not carry
physically interesting information.

As an alternative, one may adopt a more naive
approach, which consists in considering the operators
$\ln(\hat{v}_a)$ just as multiplication operators
acting on the elements of $\Hil_{v_1}$ in the
$v_a$-representation:
\begin{equation}\label{eq:lnv-def}
[\ln(\hat{v}_a)\,\Phi_{v_1}](v_2,v_3)=\
\ln(v_a)\,\Phi_{v_1}(v_2,v_3).
\end{equation}
We can rewrite the action of these operators in terms
of the variables $\omega_a$ and represent them as
operators on the physical Hilbert space $\Hilpe$ using
Eq. \eqref{eq:chidef}. In particular,
$\ln(\hat{v}_2)_{v_1}$ acts on
$\tilde\Phi\in\Sch(\re^2)\subset\Hilpe$ as follows:
\begin{equation}\label{eq:dir-logv}
\begin{split}
[&\ln(\hat{v}_2)_{v_1}\tilde\Phi](\omega_2,\omega_3)\
=\
\frac{1}{e^{\varepsilon_1}_{\omega_1}(v_1)} \\
&\times \int\rd\omega_2'\, \langle
e^{\varepsilon_2}_{\omega_2}| \ln(\hat{v}_2)\,
e^{\varepsilon_2}_{\omega_2'} \rangle_{\mathcal
H_{\varepsilon_2}^+}\,
e^{\varepsilon_1}_{\omega_1(\omega_2',\omega_3)}(v_1)\,
\tilde\Phi(\omega_2',\omega_3) ,
\end{split}
\end{equation}
whereas the action of $\ln(\hat{v}_3)$ is the same
with the subindex 2 replaced with 3. This implies
immediately that two operators at different times,
e.g. $\ln(\hat{v}_a)_{v_1}: \Hilpe\to\Hilpe$ and
$\ln(\hat{v}_a)_{v_1^\star}:\Hilpe\to\Hilpe$, are related
via the transformation
\begin{equation}\label{eq:dir-Q}\begin{split}
\widehat Q_{v_1,v_1^\star}\ :\ \Hilpe\ &\to\ \Hilpe\ , \\
[\widehat Q_{v_1,v_1^\star}\tilde\Phi](\omega_2,\omega_3)\
&=\ \left[\frac{e^{\varepsilon_1}_{\omega_1}(v_1)}
{e^{\varepsilon_1}_{\omega_1}(v_1^\star)}\right]
\tilde\Phi(\omega_2,\omega_3),
\end{split}\end{equation}
so that
\begin{equation}\label{eq:dir-lnv-rel}
\ln(\hat{v}_a)_{v_1^\star}\ =\widehat Q_{v_1,v_1^\star}
\ln(\hat{v}_a)_{v_1} \widehat Q_{v_1^\star,v_1} .
\end{equation}

Since the amplitude
$|e^{\varepsilon_1}_{\omega_1}(v_1)|$ changes
significantly both when $\omega_1$ or $v_1$
varies, the operators $\widehat Q_{v_1,v_1'}$ are
not unitary. Hence, the family of observables defined
here \emph{fails to be unitarily related}.

In order to attain a notion of nontrivial unitary
evolution in $v_1$, we propose in the next subsection
a particular construction which exploits the
asymptotic properties of
$e^{\varepsilon_1}_{\omega_1}(v_1)$ and their relation
with their WDW analogs $\ub{e}_{\omega_1}(v_1)$.

\subsection{$v_1$-observables on components}
\label{sec:v-obs}

The success of the construction of Subsec.
\ref{sec:WDWphys-evo} to provide a nontrivial
evolution picture for the WDW model rests in the form
of the eigenfunctions $\ub{e}_{\omega_1}(v_1)$ of the
operator $\ub{\widehat\Theta}_1$, which are
essentially rotating complex functions. In LQC, the
analogous eigenfunctions
$e^{\varepsilon_1}_{\omega_1}(v_1)$ oscillate rather
than rotate. Furthermore, on each of the
subsemilattices ${}^{(4)}\!\lat_{\varepsilon_1}^+$ and
${}^{(4)}\!\lat_{\varepsilon_1+2}^+$, these elements
converge to a combination of incoming and outgoing WDW
eigenfunctions, both contributing with equal
amplitude. In this sense, each eigenfunction
$e^{\varepsilon_1}_{\omega_1}(v_1)$ of
$\widehat\Theta_1$ can be interpreted as a standing
wave, which contains both components moving forward
and backward in time. This interpretation is supported
by the studies of the classical effective dynamics of
the system performed in Ref. \cite{Geff}, where a
bounce in the internal time $v_1$ is observed.

These considerations suggest that, rather than trying
to construct the analogs of the WDW observables
$\ln(\hat{v}_a)_{v_1}$, one should build instead two
separate families $\ln(\hat{v}_a)_{v_1}^\pm$, each
corresponding to one of the two commented components
of the wave function. With respect to the procedure
specified in Subsec. \ref{sec:WDWphys-evo}, this can
be viewed as a specific choice of \emph{two} (instead
of one) auxiliary Hilbert spaces: $\Hil'{}^+$ and
$\Hil'{}^-$.

In order to define the decomposition in a precise
form, we first introduce the following transformation
of the eigenfunctions of $\widehat\Theta_1$, defined
in the distributional sense on each of the
subsemilattices ${}^{(4)}\!\lat_{\varepsilon_1}^+$ and
${}^{(4)}\!\lat_{\varepsilon_1+2}^+$ separately:
\begin{equation}\label{eq:rot-dec}
e^{\tilde{\varepsilon_1}}_{\omega_1}\ \to\
e^{\tilde{\varepsilon_1}s}_{\omega_1}\ =\
{\mathcal{F}}^{-1}\theta[-s(b_1-\pi/2)]{\mathcal{F}}
e^{\tilde{\varepsilon_1}}_{\omega_1} ,
\end{equation}
where $s\in\{+,-\}$, $\tilde{\varepsilon}_1\in
\{\varepsilon_1,\varepsilon_1+2\}$, $\theta$ is a
Heaviside step function, $b_1$ is the momentum
conjugate to $v_1$ [see Eqs. \eqref{eq:b-def} and
\eqref{eq:bv-poiss}], and $\mathcal{F}$ is a discrete
Fourier transform analogous to the one defined for
isotropic systems in Refs. \cite{kl-sadj,acs}:
\begin{equation}\label{eq:fourier-def}
[{\mathcal{F}} f](b_1)\ =\
\sum_{v_1\in{}^{(4)}\!\lat_{\tilde{\varepsilon}_1}^+}
f(v_1) v_1{}^{-\frac{1}{2}}
e^{-\frac{i}{2}v_1b_1},\quad b_1\in[0,\pi].
\end{equation}
The introduction of the rescaling by
$v_1{}^{-\frac{1}{2}}$ in this transformation is
needed to develop the analysis of the evolution in
terms of $b_1$ that we carry out in Sec.
\ref{sec:lqc-b-rep}. Therefore, we use the same
Fourier transform here.

The transformation \eqref{eq:rot-dec} essentially
extracts in each subsemilattice the components
(labeled by + and -) of
$e^{\tilde{\varepsilon_1}}_{\omega_1}$ that
respectively converge, in the large $v_1$ limit, to
the WDW analogs $\ub{e}_{-|\omega_1|}$ and
$\ub{e}_{|\omega_1|}$, which move backward and forward
in time.

The functions $e^{\tilde{\varepsilon_1}s}_{\omega_1}$
sum up to the original eigenfunctions
$e^{\tilde{\varepsilon_1}}_{\omega_1}$, therefore one
can split any wave function $\Phi_{v_1}$, defined in
Eq. \eqref{eq:phi-chi}, into rotating components
$\Phi_{v_1}^{s}$ simply by replacing the
eigenfunctions $e^{\varepsilon_1}_{\omega_1}$ in Eq.
\eqref{eq:chidef} with $e^{\varepsilon_1s}_{\omega_1}$
\cite{rot-comp}. However, the Hilbert spaces
$\Hil^{s}_{v_1}\supset\Phi_{v_1}^{s}$, which are the
analogs of $\Hil_{v_1}$ [in the sense of the
definition \eqref{eq:Hv-def}], still have different
inner products for different $v_1$, since
$|e^{\varepsilon_1s}_{\omega_1}|$ depends on $v_1$.
Thus, to ``synchronize'' the norms we include one more
step in the splitting, namely the normalization of
$e^{\varepsilon_1s}_{\omega_1}$ into pure phases, and
introduce the corresponding auxiliary Hilbert spaces
$\Hil'{}^{s}$, analogs of \eqref{eq:wdw-Hprime}. The
final splitting $\Hilpe\to\Hil'{}^{s}$ is thus defined
(on the bases of the Hilbert spaces) as follows
\begin{equation}\label{eq:split-eig}
e^{\varepsilon_1}_{\omega_1}\ \mapsto\
{e'}^{\varepsilon_1s}_{\omega_1}\ :=\
|\omega_2+\omega_3|^{\frac{1}{2}}
\frac{e^{\varepsilon_1s}_{\omega_1}}{
|e^{\varepsilon_1s}_{\omega_1}|}.
\end{equation}

The implementation of this splitting allows us to
define the projection $\widehat R^{s}_{v_1}$ of the
physical states onto rotating components:
\begin{subequations}\label{eq:rotP-def}\begin{align}
&\widehat R^{s}_{v_1} : \Hilpe\to\Hil'{}^{s},\quad
\Hil'{}^{s} = L^2(\re^2,\rd\omega_2\rd\omega_3),
\notag \\
\begin{split}
&[\widehat R^{s}_{v_1}\tilde{\Phi}](\omega_2,\omega_3)
= \tilde{\boldsymbol{\chi}}^{s}_{v_1}(\omega_2,\omega_3)\\
&\hphantom{[\widehat
R^{s}_{v_1}\Phi](\omega_2,\omega_3)} :=
\tilde{\Phi}(\omega_2,\omega_3)
{e'}^{\varepsilon_1s}_{\omega_1}(v_1).
\end{split}
\tag{\ref{eq:rotP-def}}
\end{align}\end{subequations}
Using these projections we can finally define two
families of observables:
$\ln(\hat{v}_a)_{v_1}^{+}:\Hilpe\to\Hilpe$ and
$\ln(\hat{v}_a)_{v_1}^{-}:\Hilpe\to\Hilpe$, starting
from the kinematical operators $\ln(\hat{v}_a)$. Their
action on $\Hilpe$ is analogous to Eq.
\eqref{eq:dir-logv},
\begin{subequations}\label{eq:logV-def}
\begin{align}
[&\ln(\hat{v}_a)_{v_1}^{s}\tilde\Phi](\omega_2,\omega_3)\
=\ \frac{1}{{e'}^{\varepsilon_1s}_{\omega_1}(v_1)}
\tag{\ref{eq:logV-def}}\\
&\times \int\rd\omega_2'\, \langle
e^{\varepsilon_2}_{\omega_2}| \ln(\hat{v}_a)\,
e^{\varepsilon_2}_{\omega_2'}
\rangle_{\Hil_{\varepsilon_2}^+}\,
{e'}^{\varepsilon_1s}_{\omega_1(\omega_2',\omega_3)}(v_1)\,
\tilde\Phi(\omega_2',\omega_3) . \notag
\end{align}
\end{subequations}

Within each particular family labeled by $a$ and $s$,
two observables evaluated at different times $v_1$ and
$v_1^\star$ are related via the operators
\begin{equation}\label{eq:Q}\begin{split}
\widehat Q_{v_1,v_1^\star}^{s}\ :\ \Hilpe\ &\to\ \Hilpe\ , \\
[\widehat
Q_{v_1,v_1^\star}^{s}\tilde\Phi](\omega_2,\omega_3)\
&=\ \frac{{e'}^{\varepsilon_1s}_{\omega_1}(v_1)}
{{e'}^{\varepsilon_1s}_{\omega_1}(v_1^\star)}
\tilde\Phi(\omega_2,\omega_3),
\end{split}\end{equation}
in the following way
\begin{equation}\label{eq:logV-rel}
\ln(\hat{v}_a)_{v_1^\star}^{s}\ =\widehat
Q_{v_1,v_1^\star}^{s} \ln(\hat{v}_a)_{v_1}^{s} \widehat
Q^{s}_{v_1^\star,v_1}.
\end{equation}
Since, by definition \eqref{eq:split-eig},
$|{e'}^{\varepsilon_1s}_{\omega_1}(v_1)| =
|\omega_2+\omega_3|^{\frac{1}{2}}$ for all
$v_1\in\lat_{\varepsilon_1}^+$, the operators
$\widehat Q_{v_1,v_1^\star}^{s}$ are unitary and,
therefore, within each family the considered
observables are \emph{unitarily related}. We can again
extend the set formed by these families, adding the
operators $\widehat{{\Theta}}_a
|_{v_1}:={\widehat{{\Theta}}_a}$, to obtain a complete
set of observables.

Thus, the operators defined in Eq. \eqref{eq:logV-def}
provide a correct notion of unitary evolution.
However, this comes at a price. Owing to the
normalization \eqref{eq:split-eig}, the observables no
longer have a precise physical interpretation. Such
an interpretation can be recovered only
asymptotically for large $v_1$, where the rotating
components $e^{\tilde{\varepsilon}_1s}_{\omega_1}$
approach their WDW analogs (see Subsec.
\ref{sec:fr-Hph}) and
$|e^{\varepsilon_1s}_{\omega_1}|$ converges to an
$\omega_a$-independent function [see Eq.
\eqref{eq:WDW-eig}]. As a consequence, one can
interpret the operators $\ln(\hat{v}_a)_{v_1}^{s}$
only approximately as evaluating $\ln(v_a)$ at the
given value of $v_1$ on the component that is moving
forward (for negative sign) or backward (for positive
sign) in time. The approximation improves as $v_1$
increases; however, for $v_1$ of the order of
$\omega_1$ or smaller (where the effective theory
predicts a bounce in $v_1$) all the precision is lost.
An illustrative argument which shows the
``unreliability'' of the interpretation of
$\ln(\hat{v}_a)_{v_1}^{s}$ in such regime is presented
in Subsec. \ref{sec:v-results}, where we discuss the
application of the construction introduced here to
analyze the dynamics of physical states which are
semiclassical at late times.

\subsection{Numerical aspects of the analysis}
\label{sec:v-num}

In this subsection we describe the numerical methods
used to analyze the dynamics of the model. The reader
that is not interested in these numerical aspects can
safely skip this part and go to Subsec.
\ref{sec:v-results}, where the results are presented.

We focus our discussion on physical states that are
semiclassical at late times; more precisely, on
Gauss\-ian states \eqref{eq:gauss-state} peaked around
large $(\omega_2^\star,\omega_3^\star)$ [whose late
time trajectory is determined by $(\beta^2,\beta^3)$
via Eq.~\eqref{eq:limit-trajs}]. The wave functions
corresponding to such states, given by the integral
\eqref{eq:fdec}, are next evaluated applying the
trapezoid method in the domain $\omega_a \in
[\omega_a^\star-5\sigma_a,\omega_a^\star+5 \sigma_a]$.
$\tilde{\Phi}(\omega_2,\omega_3)$ has been probed
within the uniform grid defined by the split of the
domain into at least $2\omega_a^\star$ subintervals in
each direction.

To find the expectation values of $\ln(v_a)_{v_1}^{s}$
we have used the expression of the elements
$\tilde{\boldsymbol{\chi}}_{v_1}^{s}(\omega_2,\omega_3)$
of $\Hil'{}^{s}$ as functions of $v_a$,
\begin{equation}\label{eq:aux-chi}
{\boldsymbol{\chi}}_{v_1}^{s}(v_2,v_3) =
\int_{\re^2}\rd\omega_2\rd\omega_3
\tilde{\boldsymbol{\chi}}_{v_1}^{s}(\omega_2,\omega_3)
e^{\varepsilon_2}_{\omega_2}(v_2)
e^{\varepsilon_3}_{\omega_3}(v_3).
\end{equation}

Since $e^{\varepsilon_a}_{\omega_a}(v_a)$ form
orthonormal bases on their respective kinematical
spaces $\Hil_{\varepsilon_a}^+$, the physical inner
product \eqref{physpa} on $\Hil'{}^{s}$ takes a very
simple form in the $v_a$-representation
\begin{equation}\label{eq:aux-ip}
\langle{\boldsymbol{\chi}}_{v_1}^{s}|
{\boldsymbol{\chi}}_{v_1}^{\prime s}\rangle =
\sum_{\lat_{\varepsilon_2}^+\times\lat_{\varepsilon_3}^+}
\bar{{\boldsymbol{\chi}}}_{v_1}^{s}(v_2,v_3)
{\boldsymbol{\chi}}_{v_1}^{\prime s}(v_2,v_3).
\end{equation}
Furthermore, on
${\boldsymbol{\chi}}_{v_1}^{s}(v_2,v_3)$ the
observables $\ln(\hat{v}_a)_{v_1}^{s}$ act just as
multiplication operators
\begin{equation}\label{eq:aux-v-act}
[\ln(\hat{v}_a)_{v_1}^{s}
{\boldsymbol{\chi}}_{v_1}^{s}](v_2,v_3) =
\ln({v}_a){\boldsymbol{\chi}}_{v_1}^{s}(v_2,v_3).
\end{equation}
Hence, their expectation values on the state $\Phi$
are \begin{equation}\label{eq:aux-v-exp}
\langle\Phi|\ln(\hat{v}_a)_{v_1}^{s}\Phi\rangle =
\|{\boldsymbol{\chi}}_{v_1}^{s}\|^{-2}
\sum_{\lat_{\varepsilon_2}^+\times\lat_{\varepsilon_3}^+}
\ln(v_a) |{\boldsymbol{\chi}}_{v_1}^{s}(v_2,v_3)|^2 ,
\end{equation}
where ${\boldsymbol{\chi}}_{v_1}^{s}$ is related to
$\Phi$ via Eq. \eqref{eq:rotP-def}.

The dispersions corresponding to $\ln(\hat{v}_a)_{v_1}^{s}$
are given by the standard formula
\begin{equation}\label{eq:disp-def}
\langle\Delta\ln(\hat{v}_a)_{v_1}^{s}\rangle^2 =
\langle(\ln(\hat{v}_a)_{v_1}^{s})^2\rangle -
\langle\ln(\hat{v}_a)_{v_1}^{s}\rangle^2,
\end{equation}
where the expectation values of
$[\ln(\hat{v}_a)_{v_1}^{s}]^2$ are evaluated
as we have explained for
$\langle\Phi|\ln(\hat{v}_a)_{v_1}^{s}\Phi\rangle $.

In the numerical simulations, in order to calculate
these expectation values we have first evaluated the
components ${e}^{\varepsilon_1s}_{\omega_1}$ given in
Eq. \eqref{eq:rot-dec} with the use of the Fast
Fourier Transform (FFT) algorithm, and then the
normalized components
${e'}^{\varepsilon_1s}_{\omega_1}$ according to Eq.
\eqref{eq:split-eig}. After calculating the auxiliary
profiles
$\tilde{\boldsymbol{\chi}}^{s}_{v_1}(\omega_2,\omega_3)$,
defined by Eq. \eqref{eq:rotP-def}, we have evaluated
their wave functions
${\boldsymbol{\chi}}_{v_1}^{s}(v_2,v_3)$ via direct
integration of Eq. \eqref{eq:aux-chi}, in a similar
way as explained for the evaluation of
$\Phi(\vec{v})$. Next, we have found the expectation
values computing the summation in Eq.
\eqref{eq:aux-v-exp}. The domains of $v_a$ have been
restricted to
$\lat_{\varepsilon_a}^+\cap[0,4\omega_a^\star]$. The
values of $v_1$ have been chosen from the set
$\lat_{\varepsilon_1}^+\cap[0,\,2\cdot 10^5]$.
Finally, we have performed simulations for various
values of $\vec{\varepsilon}$ and for $\omega_a^\star$
ranging from $2.5\cdot 10^2$ to $10^3$.

\subsection{Results and discussion}
\label{sec:v-results}

\begin{figure*}[htb!]
\begin{center}
$(a)$\hspace{3.2in}$(b)$
\end{center}
\includegraphics[width=3.2in]{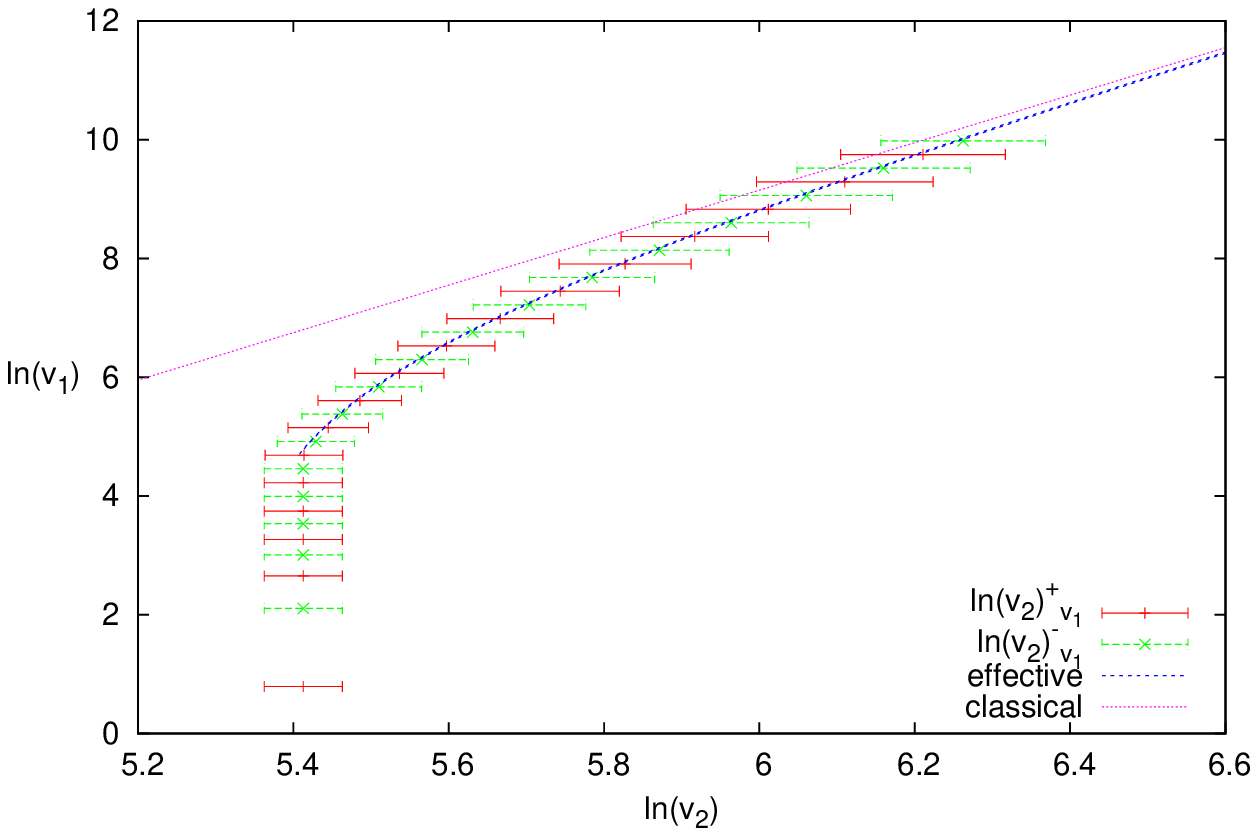}
\includegraphics[width=3.2in]{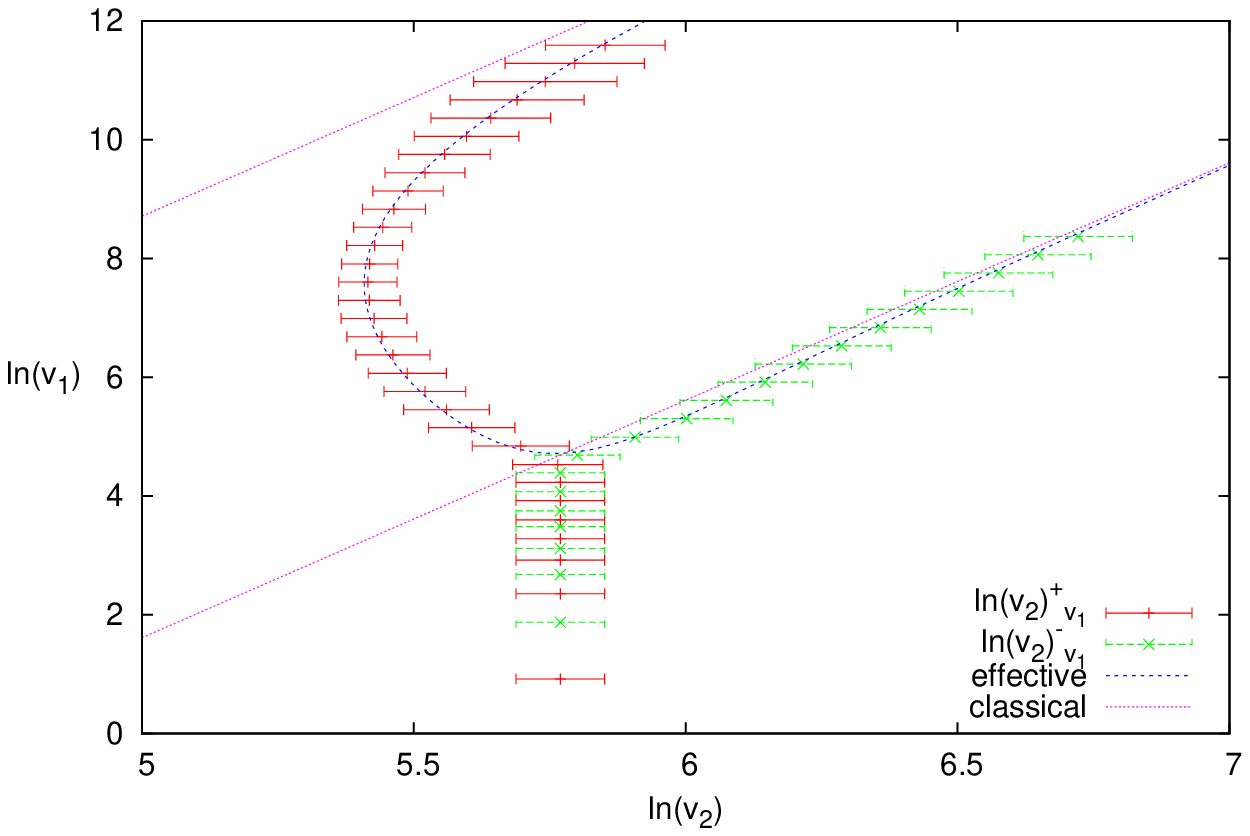}
\caption{For Gaussian states [with profile
\eqref{eq:gauss-state}], the expectation values and
dispersions of the observables
$\ln(\hat{v}_a)_{v_1}^{+}$ and
$\ln(\hat{v}_a)_{v_1}^{-}$ [corresponding to
epochs when the wave packet is moving forward (green
errorbars) and backward (red errorbars) in time,
respectively] are compared with classical (pink
lines) and effective (blue dotted line) trajectories.
The states are peaked at
$\omega^\star_2=\omega^\star_3=10^3$ with the relative
dispersions
$\Delta\Theta_2/\Theta_2=\Delta\Theta_3/\Theta_3=0.05$.
The phases are, respectively, $\beta^2=\beta^3=0$ for
$(a)$, and $\beta^2=\beta^3=0.1$ for $(b)$.
The expectation values follow the effective trajectory
till the point of bounce in $v_1$, where the dynamics
``freezes''. On the other hand away from the bounce
the quantum trajectory approaches the classical ones.
In particular, in $(a)$ the trajectories before and
after the bounce coincide on the plane $v_1 - v_2$.}
\label{fig:v-traject}
\end{figure*}

In Fig. \ref{fig:v-traject} we show two examples of
the results of our numerical study. In all cases, the
analysis reveals that
\begin{itemize}
\item Inasmuch as the expectation values
\eqref{eq:aux-v-exp} are
concerned, the states remain \emph{sharply peaked for
all} values of $v_1$. This applies to \emph{both}
components moving forward and backward in time.
\item For $v_1\gg\omega^\star_a$, the expectation values
\emph{follow the classical trajectories}, whereas,
when the universe approaches the classical
singularity, the discrete geometry effects induce
repulsive forces which cause \emph{bounces} in both
$v_2$ and $v_3$ at the values predicted by the
effective dynamics.
\end{itemize}

These results prove the robustness of the big bounce
scenario of LQC, extending its validity to another
system: the vacuum Bianchi I cosmologies. They confirm
as well the ability of the effective dynamics to
predict correct results with errors much lower than
the spread of the wave function.

Nonetheless, there is one aspect of the results which
requires special comments. As one can see in Fig.
\ref{fig:v-traject}, while for large $v_1$ the
expectation values of the observables
$\ln(\hat{v}_a)_{v_1}^{s}$ follow the classical
trajectories (with decreasing precision for decreasing
$v_1$), for $v_1\lesssim 0.2\,\omega_1(\omega_2^\star,
\omega_3^\star)$ the expectation values of both
families ``freeze'' at the same trajectory of constant
$v_2$ and $v_3$. However this behavior does not
correspond to that of $\ln(v_a)$ at given $v_1$ in any
physical way, since $\ln(\hat{v}_a)_{v_1}^{s}$ no
longer approximates it (for each $a$) in that region.

\begin{figure}
\includegraphics[width=3.2in]{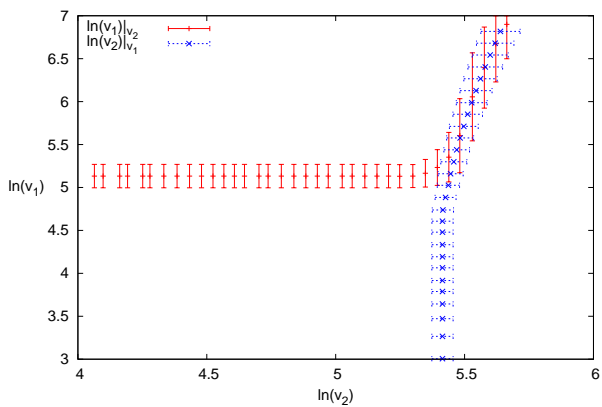}
\caption{Comparison of the expectation values of the
two families of observables $\ln(\hat{v}_2)_{v_1}^{+}$
(blue) and $\ln(\hat{v}_1)_{v_2}^{+}$ (red),
corresponding to two different choices of emergent
time, calculated on the state considered in Fig.
\ref{fig:v-traject}a.} \label{fig:v1-v2-comp}
\end{figure}

To illustrate this fact, let us consider a slight
modification of the proposed scheme for the
construction of observables. So far, we have used
$v_1$ as an emergent time; however, owing to the
symmetry of the system, any of the variables $v_i$ can
play this role. Therefore, for a given physical state
which is semiclassical at late times, we can consider,
e.g., the two following constructions:
$\ln(\hat{v}_a)_{v_1}^{s}$ and
$\ln(\hat{v}_{a'})_{v_2}^{s}$, where $a'=1,3$, and
compare the corresponding expectation values. Such
comparison is presented in Fig.~\ref{fig:v1-v2-comp}.
Outside the regions where the effective dynamics
predicts bounces, the trajectories formed by the two
considered families agree. However, once the value of
the emergent time corresponding to each of the
families drops below that of the point where the
effective bounce occurs, we observe significant
discrepancies in the resulting trajectories. Indeed,
while one family shows a bounce, the other predicts
the freezing of the trajectory, and vice versa. This
implies that in such regions the trajectories
\emph{cannot be interpreted} as corresponding to the
value of $\ln(v_i)$ at given emergent time in any
(even approximate) sense. The applicability of the
observables associated with $v_i$-slices and with
$v_i$ as an emergent time is thus limited from a
physical viewpoint.

In order to obtain a more reliable description, with a
valid physical interpretation in all the interesting
regions, we have to use a different construction of
observables. Such construction will be presented in
the next section, where as emergent time, instead of
$v_1$, we choose its conjugate momentum $b_1$.

\section{Unitary evolution on $b$-sections}\label{sec:lqc-b-rep}

One of the main reasons why the analog of the WDW
observables defined on $v_1$-slices fail to provide a
unitary evolution in the loop quantization is the fact
that the variable chosen as emergent time does not
possess all the ideal properties of this kind of time.
Indeed, an analysis of the dynamics at the effective
level \cite{Geff} and of the properties of the
eigenfunctions (see Subsec. \ref{sec:fr-Hph}) shows
that $v_1$ is not monotonous in the proper time (see
also Appendix \ref{app:eff}). This forced us to define
two families of operators corresponding to epochs when
the universe is respectively contracting and expanding
in $v_1$. Such a splitting can be avoided if we
identify a phase space variable which changes
monotonously with time. Here again the effective
dynamics comes to the aid, showing that any
canonically conjugate momentum $b_i$ of $v_i$
possesses the desired property, so that it may be a
promising candidate for an emergent time in the
genuine quantum theory.

In this section we explore this possibility by
replacing the configuration variable $v_1$ with its
momentum $b_1$ and representing $\Hilpe$ by Hilbert
spaces of data on $b_1=\const$ slices (in analogy with
the $v_1$-slices discussed in Subsec. \ref{sec:vHil}).
This step is described in Subsec. \ref{sec:b-slices}.
We then introduce in Subsec. \ref{sec:b-obs} the
observables corresponding to $\ln(v_i)$ at a given
$b_1$. The constructed family of operators is used in
Subsec. \ref{sec:b-num} to carry out a numerical
analysis of physical states, whose results are
presented in Subsec. \ref{sec:b-res}.

\subsection{$b$-representation}
\label{sec:b-slices}

Let us start by choosing at the classical level the
momentum $b_1$ conjugate to $v_1$ which is defined
through Eqs. \eqref{eq:b-def} and \eqref{eq:bv-poiss}.
In our quantization, the transformation between the
``position'' and ``momentum'' representations is given
by the formula
\begin{equation}\label{eq:lqc-v-b-trans}
[{\mathcal{F}}\psi](b_i) =
\sum_{\lat^+_{\varepsilon_i}} \psi(v_i)
|v_i|^{-\frac{1}{2}} e^{-\frac{i}{2}v_ib_i}.
\end{equation}
As commented in Subsec. \ref{sec:v-obs}, we have
introduced an additional rescaling by $|v_i|^{-1/2}$
in comparison with Eq. \eqref{eq:v-b-trans}. The
reason for that is twofold: the rescaling guarantees a
proper convergence in order to perform the
transformation numerically and, in addition, ensures a
more convenient behavior of the transformed
eigenfunctions (see Subsec. \ref{sec:b-num}). Under
this transformation, the kinematical operators which
enter the Hamiltonian constraint and the observables
change as follows
\begin{eqnarray}\label{eq:lqc-ops-b}
& \hat{v_i}  \to  2i\partial_{b_i},  \nonumber\\
&  |\hat{v}_1|^{1/2}\left(\hat{\mathcal
N}_{2\bar\mu_i} - \hat{\mathcal
N}_{-2\bar\mu_i}\right) |\hat{v}_1|^{-1/2} \to
2i\sin(\hat{b}_i),
\end{eqnarray} where
$\hat{b}_i$ acts in the new representation by
multiplication.

Since, in order to construct the physical state, we
restrict the analysis to one superselection sector,
with functions supported on $\lat_{\varepsilon_i}^+$,
the same restriction applies to the transformation
\eqref{eq:lqc-v-b-trans}. In this case, the domain of
$b_i$ can be taken as the unit circle $b_i\in S^1$.
Moreover, to further adapt the transformation to our
specific system, let us note the following:
\begin{enumerate}[(i)]
\item The restrictions of the eigenfunctions of
$\widehat\Theta_i$ (supported on
$\lat_{\varepsilon_i}^+$) to the subsemilattices
${}^{(4)}\lat_{\varepsilon_i}^+$ and
${}^{(4)}\lat_{\varepsilon_i+2}^+$ are eigenfunctions
of $\widehat\Theta_i^2$, which differs from the
evolution operator of the isotropic system just by a
compact operator (see Ref. \cite{mmp} and Subsec.
\ref{sec:fr-Hph}).\label{it:split}
\item
Since, e.g., any $v_1={\rm const}$ section contains
all the information about a physical state, in
particular, the restriction of a physical state to one
of the subsemilattices for $v_1$ uniquely determines
the other. \label{it:subsemi}
\item In the isotropic model and within the context of
solvable LQC (sLQC) \cite{acs} a similar
transformation allows one to select the domain of the
corresponding momentum $b$ as $[0,\pi)$. Furthermore,
the sLQC analog of $\widehat\Theta^2_i$ is
proportional to $[\sin(b)\partial_b]^2$. This in turn
allows one to write the constraint as a Klein-Gordon
equation. \label{it:KG}
\end{enumerate}

This last point indicates that it is convenient to
apply the transformation \eqref{eq:lqc-v-b-trans} to
functions supported on
${}^{(4)}\lat_{\varepsilon_i}^+$ and
${}^{(4)}\lat_{\varepsilon_i+2}^+$ independently. The
domain of $b_i$ consists then of two copies of the
circle with radius one half (i.e., the periodicity of
$b_i$ is $\pi$ in each copy). As a consequence, one
can define the transformation $\mathcal{F}$ such that
\begin{equation}
[{\mathcal{F}}_1\psi](b_i) = \sum_{v_i\in
{}^{(4)}\lat_{\varepsilon_i}^+} \psi(v_i)
v_i{}^{-\frac{1}{2}} e^{-\frac{i}{2}v_ib_i},
\end{equation}
and, analogously, the transformation ${\mathcal{F}}_2$
by choosing the summation over
${}^{(4)}\lat_{\varepsilon_i+2}^+$ instead of
${}^{(4)}\lat_{\varepsilon_i}^+$.

If we transform the eigenfunctions of
$\widehat\Theta_1$, we can define a ``mixed''
representation of physical states
\begin{equation}\begin{split}
\Phi(b_1,v_2,v_3) &=
\int_{\re^2}\rd\omega_2\rd\omega_3
\tilde{\Phi}(\omega_2,\omega_3)
\tilde{e}^{\varepsilon_1}_{\omega_1}(b_1) \\
&\times  e^{\varepsilon_2}_{\omega_2}(v_2)
e^{\varepsilon_3}_{\omega_3}(v_3) ,
\end{split}\end{equation}
where $\tilde{e}^{\varepsilon_1}_{\omega_1}(b_1)$ is
\begin{equation}\label{eq:e-brep-def}
\tilde{e}^{\varepsilon_1}_{\omega_1}(b_1) =
\begin{cases}
[{\mathcal{F}}_1 e^{\varepsilon_1}_{\omega_1}](b_1),
&  b_1 \in [0,\pi) \\
[{\mathcal{F}}_2
e^{\varepsilon_1}_{\omega_1}](b_1-\pi),  &  b_1 \in
[\pi,2\pi),
\end{cases}
\end{equation}
As discussed above, we restrict our considerations,
for example, to the domain of $b_1$ which corresponds
to the first of these cases: $b_1 \in [0,\pi)$. Within
this domain, one can now apply the analog of the
constructions presented in Sec. \ref{sec:wdw-analog}
and Sec. \ref{sec:v-evo}. In doing so, we introduce
the counterpart of the spaces $\Hil_{v_1}$ [given in
Eq. \eqref{eq:Hv-def}], which are defined on surfaces
of constant $b_1$,
\begin{equation}
\Hil_{b_1} := L^2(\re^2, |\omega_2+\omega_3|
|\tilde{e}^{\varepsilon_1}_{\omega_1}(b_1)|^{-2}
\rd\omega_2\rd\omega_3),
\end{equation}
and the transformation of the physical states into
elements in these spaces
\begin{subequations}\begin{align}
\widehat P_{b_1}:\Hilpe\ &\to\Hil_{b_1}, \\
\begin{split}
[\widehat P_{b_1}\tilde{\Phi}](\omega_2,\omega_3)\ &=\
\tilde{\Phi}_{b_1}(\omega_2,\omega_3) \\
&:=\ \tilde{e}^{\varepsilon_1}_{\omega_1}(b_1)
\tilde{\Phi}(\omega_2,\omega_3).
\end{split}
\end{align}\end{subequations}
Each function $\tilde{\Phi}_{b_1}$ at any given slice
$b_1=\const$ determines the physical state uniquely,
thus it can be understood as the initial data at that
slice. Therefore, one may develop a notion of
evolution in $b_1$ using the corresponding map between
initial data spaces. It is worth noticing,
nonetheless, that, unlike in sLQC (and with the chosen
factor ordering), the Hamiltonian constraint \eqref{C}
in a $b_i$-representation cannot be expressed as a
differential equation of finite order.

\subsection{Observables associated with $b$-slices}\label{sec:b-obs}

With the spaces of initial data at constant $b_1$ at
our disposal, we can now define the observables
$\ln(v_i)_{b_1}$. For $i=2,3$ the construction is
completely parallel to that presented in Subsec.
\ref{sec:pre-v-obs}; however, it suffers from a
similar problem of lack of a unitary relation between
the operators in each family of observables. Here,
nevertheless, the problem is much less severe. Indeed,
the functions
$\tilde{e}^{\varepsilon_1}_{\omega_1}(b_1)$ are
(except possibly for small $\omega_1$) rotating
functions that never vanish (see Subsec.
\ref{sec:b-num}). This property has a qualitative
analytical explanation. Actually, the operator
${\mathcal{F}}_1(\widehat\Theta_1^2)$ is, up to a
compact correction, equal to $-[12\pi \gamma
G\sin(b_1)\partial_{b_1}]^2$, whose eigenspaces are
spanned by basis elements of the form $N e^{\pm
ikx(b_1)}$, with $x(b_1) := \ln[\tan(b_1/2)]$ and
where $N=N(k)$ is a normalization factor. Furthermore,
once we restrict the space to functions supported on
$v_1>0$ (as happens to be the case in our model), the
eigenfunctions reduce just to purely rotating
functions [with vanishing contribution of one of the
two phases $\pm ikx(b_1)$]. These eigenfunctions will
differ from those of the actual operator
${\mathcal{F}}_1(\widehat\Theta_1^2)$ just by small
corrections whose (kinematical) norm decreases with
$\omega_1$.

As a consequence, we do not need to split the
eigenfunctions into rotating components in order to
form a family of unitarily related observables.
The following parallel of the transformation
\eqref{eq:split-eig}:
\begin{equation}\label{eq:b-norm}
\tilde{e}^{\varepsilon_1}_{\omega_1}(b)\ \to\
\tilde{e}'{}^{\varepsilon_1}_{\omega_1}(b) =
|\omega_2+\omega_3|^{\frac{1}{2}}
\frac{\tilde{e}^{\varepsilon_1}_{\omega_1}(b)}
{|\tilde{e}^{\varepsilon_1}_{\omega_1}(b)|}
\end{equation}
is sufficient to build analogs $\widehat R_{b_1}$ of
the operators $\widehat R^{s}_{v_1}$,
\begin{subequations}\begin{align}
\widehat R_{b_1}:\Hilpe &\to \Hil'_{b_1}:=
L^2(\re^2,\rd\omega_2\rd\omega_3)\ , \\
[\widehat R_{b_1}\tilde{\Phi}](\omega_2,\omega_3)&=
\tilde{{\boldsymbol{\chi}}}_{b_1}(\omega_2,\omega_3)
\\
&:= \tilde{\Phi}(\omega_2,\omega_3)
\tilde{e}'{}^{\varepsilon_1}_{\omega_1}(b_1).
\nonumber
\end{align}\end{subequations}

The observables $\ln(\hat{v}_i)_{b_1}$ are defined as
follows. For $i=2,3$ we simply act on
\begin{equation}\label{eq:aux-chi-b}
{\boldsymbol{\chi}}_{b_1}(v_2,v_3) =
\int_{\re^2}\rd\omega_2\rd\omega_3
\tilde{{\boldsymbol{\chi}}}_{b_1}(\omega_2,\omega_3)
e^{\varepsilon_2}_{\omega_2}(v_2)
e^{\varepsilon_3}_{\omega_3}(v_3).
\end{equation}
with the kinematical operator $\ln(\hat{v}_i)$. Thus,
the action of $\ln(\hat{v}_2)_{b_1}$ on the elements
of $\Hilpe$ turns out to be
\begin{subequations}\label{eq:lnv-b-def}\begin{align}
&[\ln(\hat{v}_2)_{b_1}\tilde\Phi](\omega_2,\omega_3) =
\frac{1}{\tilde{e}'{}^{\varepsilon_1}_{\omega_1}(b_1)}
\tag{\ref{eq:lnv-b-def}}\\
&\hphantom{.}\ \times \int\rd\omega'_2 \langle
e^{\varepsilon_2}_{\omega_2}|\ln(\hat{v}_2)
e^{\varepsilon_2}_{\omega'_2}
\rangle_{\Hil_{\varepsilon_2}^+}\,
\tilde{e}'{}^{\varepsilon_1}_{\omega_1(\omega'_2,\omega_3)}
(b_1)\tilde\Phi(\omega'_2,\omega_3), \notag
\end{align}\end{subequations}
and similarly for $i=3$. These families are unitarily
related by the very same reason explained for
$\ln(\hat{v}_a)_{v_1}$ in Subsec. \ref{sec:v-obs}, and
can be extended equally to a complete set of
observables.

As an aside, let us comment that in the sLQC approach
the analog of the basis eigenfunctions
$\tilde{e}^{\varepsilon_1}_{\omega_1}(b_1)$ are pure
phases, and then the transformation \eqref{eq:b-norm}
reduces just to a unitary rescaling by
$|\omega_2+\omega_3|^{\frac{1}{2}}$.

The construction of $\ln(\hat{v}_1)_{b_1}$ at a
heuristic level is based again on the action of the
kinematical operator on some suitable modification of
the initial data at $b_1$. To give a precise
definition, we first need to express the operator
$\ln(\hat{v}_1)_{b_1}$ in the $b_1$-representation.

Note that the function $\ln(v_1)$ can be expanded
around a point $v_1^o>0$ as
\begin{equation}
\ln(v_1) = \ln(v_1^o) - \sum_{n=1}^{\infty} \frac{1}{n
(v_1^o)^n}(v_1^o-v_1)^n
\end{equation}
Promoting the terms $v_1^n$ to operators and applying
relation \eqref{eq:lqc-ops-b}, we can represent the
operator $\ln(\hat{v}_1)$ as
\begin{equation}\label{eq:lnv-brep}
\ln(\hat{v}_1) = \ln(v_1^o) - \sum_{n=1}^{\infty}
\frac{1}{n (v_1^o)^n}(v_1^o-2i\partial_{b_1})^n,
\end{equation}
defined on elements
${\boldsymbol{\chi}}_{b_1}(v_2,v_3)$ of the
kinematical Hilbert space. Changing to the
$\omega_a$-representation via Eq.
\eqref{eq:aux-chi-b}, the action on
$\tilde{\boldsymbol{\chi}}_{b_1}(\omega_2,\omega_3)$
can be expressed as
\begin{equation}
[\ln(\hat{v}_1)\tilde{{\boldsymbol{\chi}}}_{b_1}]
(\omega_2,\omega_3) = \tilde{\Phi}(\omega_2,\omega_3)
[\ln(\hat{v}_1)
\tilde{e}'{}^{\varepsilon_1}_{\omega_1}](b_1),
\end{equation}
and thus the final form of these observables reads:
\begin{eqnarray}
&\ln(\hat{v}_1)_{b_1}:\Hilpe\to\Hilpe, \nonumber\\
\label{eq:lnv-action}
&[\ln(\hat{v}_1)_{b_1}\tilde{\Phi}](\omega_2,\omega_3)
=\frac{[\ln({\hat v_1})
\tilde{e}'{}^{\varepsilon_1}_{\omega_1}](b_1)}%
{\tilde{e}'{}^{\varepsilon_1}_{\omega_1}(b_1)}
\tilde{\Phi}(\omega_2,\omega_3),
\end{eqnarray}
with $\ln({\hat v_1})$ given in Eq. \eqref{eq:lnv-brep}.

One can check by inspection that these observables are
not related by a unitary operator $\widehat
Q_{b_1,b_1^\star}: \Hilpe \to \Hilpe$, analogous to
the operators defined in Eq. \eqref{eq:Q}. Therefore,
the observables $\ln(\hat{v}_1)_{b_1}$ can play only
an auxiliary role with respect to the families of
unitarily related observables $\ln(\hat{v}_a)_{b_1}$.
In the next section we employ these families of
observables to analyze the dynamics of physical states
which are semiclassical at late times.

\subsection{Numerical analysis}\label{sec:b-num}

Let us briefly discuss the numerical methods used in
our calculations. This is a technical subsection that
can be safely skipped if the reader is not interested
in the numerics.

As in Sec. \ref{sec:v-evo}, we concentrate our study
on the Gaussian states \eqref{eq:gauss-state}. In most
cases, the calculations are exactly the same as those
of Subsec. \ref{sec:v-num} for the observables
associated with $v_1$-slices. In particular, to
compute the expectation values and dispersions of
$\ln(\hat{v}_a)$ we have used the analogs of Eqs.
\eqref{eq:aux-chi}-\eqref{eq:disp-def}, replacing
${\boldsymbol{\chi}}_{v_1}^s$ with
${\boldsymbol{\chi}}_{b_1}$. The eigenfunctions
$\tilde{e}'{}^{\varepsilon_1}_{\omega_1}$, needed to
determine $\tilde{\boldsymbol{\chi}}_{b_1}$, have been
computed via Eq. \eqref{eq:e-brep-def} by means of a
FFT algorithm. In comparison with Subsec.
\ref{sec:v-num}, the only relevant difference appears
in the calculation of
$\langle\ln(\hat{v}_1)_{b_1}\rangle$ and
$\langle\Delta\ln(\hat{v}_1)_{b_1}\rangle$, which is
performed as follows.

In a first step, we calculate the action of
$\ln(\hat{v}_1)_{b_1}$ on $\tilde{\Phi}$ using Eq.
\eqref{eq:lnv-action}. To find the values of
$\ln(\hat{v}_1)\tilde{e}'{}^{\varepsilon_1}_{\omega_1}$,
we act with $\ln(\hat{v}_1)$ in the
$v_1$-representation, and transform the result to the
$b_1$-representation via Eq. \eqref{eq:lqc-v-b-trans}
by applying a FFT.

At this stage, however, it is necessary to comment on
a technical subtlety. In order to calculate the
desired result exactly as specified in the previous
subsection, one has to normalize the eigenfunction
according to Eq. \eqref{eq:b-norm} before acting on it
with $\ln(\hat{v}_1)$. Since the normalization must be
done in the $b_1$-representation and the action of the
operator is known instead in the $v_1$-representation,
one would have to carry out a sequence of operations,
namely, a Fourier transform, a normalization, and an
inverse Fourier transform. Unfortunately, the FFT
algorithm used for this assumes that both the function
that is transformed and the result are supported on
points which are uniformly distributed in the circle.
Since, in our case, the support of the eigenfunction
is an entire subsemilattice,
${}^{(4)}\lat_{\tilde{\varepsilon}_1}^+$, one should
restrict it to a set of the form
${}^{(4)}\lat_{\tilde{\varepsilon}_1}^+\cap [0,v_{\rm
max}]$ (where $v_{\rm max}$ has some large value),
thus removing the tail, which decays (up to
logarithmic factors) as $v_1^{-1}$. The restriction
would give rise to numerical errors which would
manifest themselves near $b_1=0$ and $b_1=\pi$. In
turn, this would lead to errors in normalization,
which would be amplified through the commented
sequence of operations (FFT $\to$ Normalization $\to$
FFT), making the results in the large volume regime
unreliable.

In order to avoid this problem, we implement a
slightly different algorithm. Instead of normalizing
$\tilde{e}^{\varepsilon_1}_{\omega_1}$, we first act
with $\ln(\hat{v}_1)$ on the nonnormalized
eigenfunctions, carrying out the normalization
afterwards. That is,
\begin{equation}\label{eq:num-act}
\tilde{e}^{\varepsilon_1}_{\omega_1}(b_1) \mapsto
\tilde{f}^{\varepsilon_1}_{\omega_1}(b_1) :=
\frac{[\ln(\hat{v}_1)
\tilde{e}^{\varepsilon_1}_{\omega_1}](b_1)}
{|\tilde{e}^{\varepsilon_1}_{\omega_1}(b_1)|}.
\end{equation}
Such a method introduces an error owing to the
difference between the normalized and nonnormalized
functions. However, one can argue that the error
can be neglected for semiclassical states peaked
around large $\omega_a^\star$, using that
$\tilde{e}'{}^{\varepsilon_1}_{\omega_1}(b_1)$ can be
approximated by its sLQC counterpart in that regime:
\begin{equation}\label{eq:b-conv}\begin{split}
\tilde{e}'{}^{\varepsilon_1}_{\omega_1}(b_1) &\sim
|\omega_2+\omega_3|^{\frac{1}{2}}
\left[ e^{i\omega_1 x(b_1)} + O(\omega_1^{-2}) \right] ,
\\
x(b) &:=
\ln{\left[\tan{\left(\frac{b}{2}\right)}\right]}.
\end{split}\end{equation}

Once the functions
$\tilde{f}^{\varepsilon_1}_{\omega_1}$ are known, we
evaluate the profiles
\begin{subequations}\label{eq:num-lnv1}\begin{align}
[\ln(\hat{v}_1){\boldsymbol{\chi}}_{b_1}](v_2,v_3) &=
\int_{\re^2}\rd\omega_2\rd\omega_3
\tilde{\Phi}(\omega_2,\omega_3)
\tilde{f}^{\varepsilon_1}_{\omega_1}(b_1) \notag \\
&\hphantom{=\int}\times
e^{\varepsilon_2}_{\omega_2}(v_2)
e^{\varepsilon_3}_{\omega_3}(v_3),
\tag{\ref{eq:num-lnv1}}
\end{align}\end{subequations}
integrating in practice the righthand side over the
intervals $\omega_a \in
[\omega_a^\star-5\sigma_a^\star,\omega_a^\star+
5\sigma_a^\star]$ via the trapezoid method.

The computed profiles are finally used to calculate
the expectation values
\begin{subequations}\label{eq:lnv1-exp}\begin{align}
\langle\Phi|\ln(\hat{v}_1)_{b_1}\Phi\rangle &=
\|{\boldsymbol{\chi}}_{b_1}\|^{-2}
\tag{\ref{eq:lnv1-exp}} \\
&\times \sum_{\bar{\lat}^{2}}
\bar{\boldsymbol{\chi}}_{b_1}(v_2,v_3)
[\ln(\hat{v}_1){\boldsymbol{\chi}}_{b_1}](v_2,v_3)
,\notag
\end{align}\end{subequations}
where owing to technical limitations the summation was
restricted to $\bar{\lat}^{2}:=(\lat_{\varepsilon_2}^+
\cap[0,4\omega_2^\star])\times
(\lat_{\varepsilon_3}^+\cap[0,4\omega_3^\star])$.

The dispersions of these observables are found using
the standard relations [similar to Eq.
\eqref{eq:disp-def}]. The expectation values
$\langle\Phi|\ln(\hat{v}_1)^2_{b_1}\Phi\rangle$ are
computed with an algorithm that is fully analogous to
the one used for
$\langle\Phi|\ln(\hat{v}_1)_{b_1}\Phi\rangle$.

In our numerical simulations, the number of points
selected for the FFT of the eigenfunctions is equal to
$2^{17}$, distributed uniformly in the set
$b_1\in[0,\pi)$. The simulations have been performed
for the same range of parameters as in Subsec.
\ref{sec:v-num}, that is, for $\omega_a^\star$ ranging
from $2.5\cdot 10^2$ to $10^3$, and for relative
dispersions in $\omega_a$ between $0.05$ and $0.1$.
The results of these simulations are discussed in the
next subsection.

\subsection{Results}
\label{sec:b-res}

\begin{figure*}[htb!]
\begin{center}
$(a)$\hspace{3.2in}$(b)$
\end{center}
\includegraphics[width=3.2in]{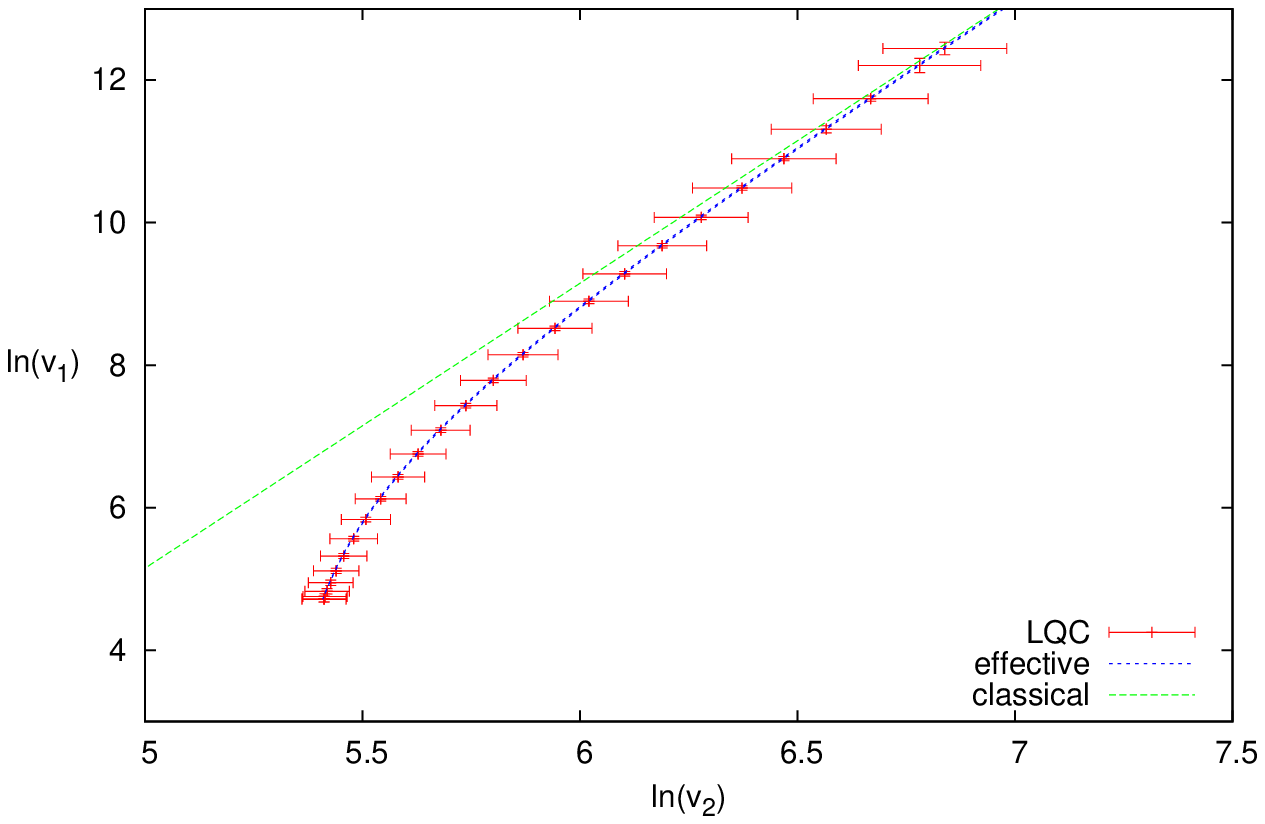}
\includegraphics[width=3.2in]{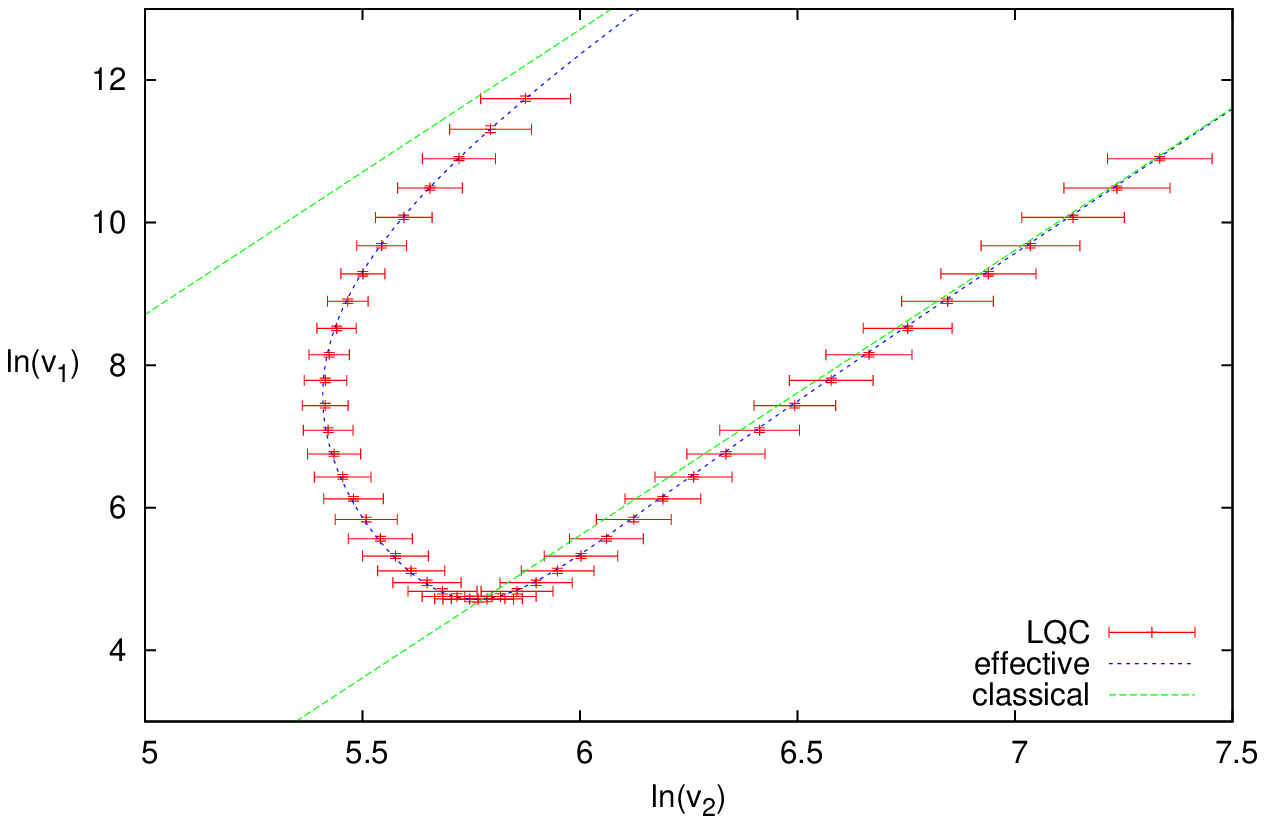}
\caption{Expectation values and dispersions of the
observables $\ln(\hat{v}_i)_{b_1}$ on the Gaussian
states [with profile \eqref{eq:gauss-state}]
presented in the $v_2-v_1$ plane and compared with
classical (green lines) and effective (blue dotted
line) trajectories.
The states are peaked at
$\omega^\star_2=\omega^\star_3=10^3$ with relative
dispersions
$\Delta\Theta_2/\Theta_2=\Delta\Theta_3/\Theta_3=
0.05$. The phases are, respectively,
$\beta^2=\beta^3=0$ for $(a)$ and
$\beta^2=\beta^3=0.1$ for $(b)$.
The expectation values follow the effective trajectory
through all the evolution. In particular, the
trajectories before and after the bounce coincide in
$(a)$.} \label{fig:b-traject}
\end{figure*}

A representative example of the numerical analysis of
the Gaussian states is presented in
Fig.~\ref{fig:b-traject}. The general conclusions of
this analysis can be summarized as follows:
\begin{itemize}
\item The states that are sharply peaked at some initial
$b_1$ remain so during the whole evolution.
\item For small $b_1$, the expectation values of the
three families of observables $\ln(\hat{v}_i)_{b_1}$
follow classical trajectories corresponding to a
universe contracting in the three directions. As $b_1$
increases and the values of $v_i$ decrease becoming of
the order of $\omega_i^\star$, we observe a deviation
from the classical dynamics which results in
independent bounces in \emph{all the three
directions}. After each bounce (in direction $i$), the
value of $v_i$ starts growing, and its dynamics
quickly approaches that of a classical expanding
universe.
\item Through the entire evolution, the genuine quantum
trajectories \emph{agree to a good precision} (less
than $10\%$ of dispersion) with those corresponding to
the classical effective dynamics presented in
Appendix~\ref{app:eff}.
\end{itemize}

The results listed above provide an additional
(independent) confirmation of the conclusions
presented in Subsec. \ref{sec:v-results}. However, the
current analysis constitutes a significant improvement with
respect to that of Sec. \ref{sec:v-evo}. Namely, while
it is still only in an asymptotic sense that the
observables used here possess a well defined physical
interpretation, the accuracy of this
interpretation remains fairly good during the entire
evolution. Unlike the situation found for
$\ln(\hat{v}_a)_{v_1}$, where the observables
completely lose a meaningful physical interpretation
in some epochs of the evolution, the level of
confidence depends now just on the peak values
$\omega^\star_a$. Furthermore, this level
improves as one considers more and more macroscopic
universes (larger $\omega_a^\star$).

\section{Conclusions and discussion}\label{sec:results}

In this work we have investigated the concept of
evolution in LQC. Our main goal has been the
construction of a well defined unitary evolution
framework with no reference to a matter field as a
clock. We have achieved this by building certain
families of unitarily related partial observables,
parameterized by geometry degrees of freedom. As a
test bed for our study, we have chosen the specific
example of a vacuum Bianchi I toroidal ($T^3$)
universe.

Within this model, we have proposed two different
constructions of observables, built out of the
kinematical operators $\ln(\hat{v}_a)$ with $a=2,3$,
and parameterized respectively by: $(i)$ the affine
parameter $v_1$ corresponding to one of the triad
coefficients $p_1$, which plays the role of a
configuration variable, $(ii)$ its conjugate momentum
$b_1$. In this way, the role of internal time has been
assigned to geometry variables, making its definition
independent of the matter content.

In order to develop and test our constructions in an
analytically solvable setting, as well as to identify
the imprint of the loop quantization in the dynamics,
we have first implemented them in the quantum version
of the system obtained via the standard methods of
geometrodynamics, i.e., in the WDW analog of our LQC
model. In both of the LQC and WDW theories, the
physical Hilbert space of our model is
$L^2\left({\re^2},|\omega_2+\omega_3|\rd\omega_2
\rd\omega_3\right)$. The difference between them is
captured in the form of the observables.

In the WDW theory the basis elements of the
kinematical Hilbert space are, essentially, rotating
functions, and owing to this property the unitarity in
the evolution is easily achieved both for cases (i)
and (ii). Furthermore, all the constructed observables
have a neat physical interpretation as the evaluation
of $\ln(v_a)$ at a given moment of $v_1$ or $b_1$,
respectively.

The constructions described above have been used to
investigate the dynamics of states which are
semiclassical at late times, with a sharply peaked
Gaussian profile. We have shown analytically that any
such state follows the classical trajectory right to
the classical singularities, which are thus not
resolved. Actually, this lack of singularity
resolution has been shown to be a general property of
the physical states, not only of the Gaussian ones.

In the case of an LQC model, the situation is more
complicated. While in case $(ii)$ we have succeeded in
applying the construction tested in the WDW model
almost straightforwardly, case $(i)$ has posed a
considerable challenge. Both a direct application or a
straightforward generalization of the scheme
introduced for WDW have resulted in operators which
either fail to provide a unitarily related family of
observables, or do not encode any physically
interesting information about the system.

To overcome this problem, we have used our knowledge
of the WDW limit of an LQC state, observing that any
limit of this kind contains two types of components,
which respectively move forward and backward in the
evolution parameter. This feature has motivated the
introduction of a similar splitting in rotating
components of the exact LQC wave functions, which has
been defined in a precise manner exploiting the
properties of the Fourier transform. The scheme for
the construction of observables defined in the WDW
case has been applied to each of the components
separately and has allowed us to attain the required
unitarity in the evolution.

In contrast to the WDW case, where the constructed
observables admitted a neat physical interpretation,
in LQC, owing to the modifications in the construction
necessary to ensure unitarity, the physical
interpretation of the operators is clear only in
approximate sense. In particular, for case $(i)$ the
interpretation of the operators as corresponding to
the value of $\ln(v_a)$ at a given $v_1$ is valid only
in the large $v_1$ limit, and accurate only for part
of the evolution. For Gaussian states peaked
around some $\omega_a^\star$, the interpretation
ceases to be acceptable when $v_1\lesssim 0.2\,
|\omega_1(\omega_2^\star,\omega_3^\star)|$, a feature
that we have seen by exchanging the spatial indexes of
the kinematical operators and of the internal time
$v_1$. Nonetheless in case (ii), while the
interpretation is still approximate, its accuracy does
not change during evolution (i.e., it does not depend
on the value of $b_1$) and is pretty good for the
physically interesting states (peaked at large
$\omega_a^\star$).

Both constructions, $(i)$ and $(ii)$, have been
employed to analyze the dynamics of Gaussian states
sharply peaked around large $\omega_a^\star$. In case
$(ii)$, and in order to bring the analysis to a common
setting ($v_1$-$v_a$ plane) to facilitate the
comparison with $(i)$, an additional family of
observables --corresponding to $\ln(v_1)$ at given
$b_1$-- has been constructed. However, these
observables are not unitarily related; therefore they
play only an auxiliary role in the discussion. The
numerical analysis reveals the following picture
(valid both for case $(i)$ and $(ii)$, unless
otherwise stated):
\begin{itemize}
\item The considered Gaussian states remain sharply
peaked (as far as the relative dispersions of the
constructed observables are concerned) through all the
evolution.
\item For large
$v_1/\omega_1(\omega_2^\star,\omega_3^\star)$, the
expectation values of the observables follow classical
trajectories, while for other values of this ratio we
have observed deviations which ultimately lead to
bounces in $v_a$.
\item Since the physical interpretation
of the observables in case $(i)$ is not reliable for
small $v_1$, because
$v_1/\omega_1(\omega_2^\star,\omega_3^\star)$ becomes
small, we have also performed the analysis in terms of
the family of observables of case $(ii)$. This
analysis has shown as well the bounce in $v_1$.
\item The expectation values of the introduced observables
(calculated in the domains where their physical
interpretation is valid) agree with the trajectories
predicted by the classical effective dynamics
presented in Appendix~\ref{app:eff}.
\end{itemize}
As a consequence, in LQC the singularity, already
resolved at a kinematical level, is also resolved
dynamically. Furthermore, the above observations imply
that, similarly to what occurs in the isotropic
system, large semiclassical universes expanding in
terms of the coefficients $p_i$ are connected to also
large and semiclassical contracting ones via a
sequence of bounces which are independent for each
direction.

Let us point out that the complication of the
numerical problems that one has to solve in the
analysis of the vacuum Bianchi I model has not allowed
us to probe states as semiclassical as the ones
studied in the isotropic case (considered for example
in Ref. \cite{aps-imp}), nor check really the long
term evolution of the system. Besides, the numerical
analysis has been restricted just to Gaussian states.
This analysis alone is not sufficient to conclude that
all states that are semiclassical before the bounces
stay so after them. However, the semianalytical study
of the WDW limit that we have performed for case $(i)$
has shown that, whenever the state has finite
expectation values and dispersions for the set of
operators \eqref{eq:loc-set} [and finite expectation
values for \eqref{eq:fin-set}], the relative
dispersions of $\ln(v_a)_{v_1}$ in both branches of
the evolution are asymptotically equal.

In summary, taking the model of the vacuum Bianchi I
universe as an example, we have constructed a notion
of unitary evolution in a well defined way and without
any reference to a matter field as a clock. We have
achieved this goal by building several families of
unitarily related observables, which in turn have
allowed us to thoroughly analyze the dynamics of the
considered model. The results of this analysis confirm
at a genuinely quantum level the robustness of the
bounce picture, as well as the preservation of the
semiclassicality across the bounces.

To conclude, we would like to comment that the
approach presented here pioneers the development of a
new methodology, which can be employed whenever one
has to analyze the evolution in absence of degrees of
freedom that are quantized in a standard way.
Therefore, it can be applied in a large variety of
polymerically quantized systems. In particular, it is
applicable to the quantization carried out in Ref.
\cite{awe-b1} where, for the analogs of the
constructions $(i)$ and $(ii)$, we can use the total
volume $v$ defined in that reference and its conjugate
momentum $b$, respectively. Nonetheless, the
prescription for the so-called improved dynamics
chosen for our study here offers us a precise control
over the introduced constructions, inasmuch as the
mathematical structure of the system has already been
carefully studied in the literature and is known in
great detail. Furthermore, it allows to test
explicitly the reliability of the results provided by
the observables. In particular, in the case of the
operators parameterized by $v_1$, we have been able to
recognize a loss of predictability (at least in terms
of a conventional physical interpretation of the
observables) by interchanging the roles of internal
time and dynamical variable between $v_1$ and $v_a$
($a=2$ or 3), a possibility which is not available in
the prescription followed in Ref. \cite{awe-b1}.

\section*{Acknowledgments}

This work was supported by the Spanish MICINN Project
FIS2008-06078-C03-03 and the Consolider-Inge\-nio 2010
Program CPAN (CSD2007-00042). The authors are
grateful to W. Kami\'nski, Javier Olmedo, Parampreet
Singh, and Jos\'e M. Velhinho for discussions.
M.M.-B. and T.P. acknowledge financial aid by the I3P
Program of CSIC and the European Social Fund, in the
case of M.M.-B. under the grant I3P-BPD2006. T.P. also
acknowledges financial support by the Foundation for
Polish Science under the grant Master and thanks the
Perimeter Institute and the Institute of Theoretical
Physics of Warsaw University for the hospitality in
the period of development of this work. M.M.-B. wants
to thank A.E.I. for warm hospitality during part of
the period when this work was done.

\appendix

\section{Dispersion in the Wheeler-DeWitt limit:
ensemble of sectors and further
considerations}\label{app:unidis}

In this appendix, we analyze the behavior of the
dispersions in the WDW regime taking into account the
eight sectors in which the LQC wave functions have a
well defined limit, rather than considering each of
these sectors independently. We want to define the ensemble of
those sectors in such a way that the properties of the expectation values
and dispersions of the LQC observables are reflected by features of their
analogs on
that ensemble. To achieve this goal, we note the following properties of the
operators
$\ln(\hat{v}_a)_{v_1}$:
\begin{enumerate}[(a)]
  \item Since each operator is defined for a definite value of $v_1$, which
    belongs just to one of the subsemilattices, ${}^{(4)}\lat_{\varepsilon_1}$
or
    ${}^{(4)}\lat_{\varepsilon_1+2}$, we can consider the states supported on
them separately. As a consequence,
    we can safely consider two independent limits corresponding to the splitting
of the support in $v_1$.
  \item For given $v_1$, the action of the operator $\ln(\hat{v}_a)_{v_1}$ can
be represented as the action of the sum of the respective restrictions to each
of the four sectors obtained with the splitting into subsemilattices in the
directions $a=2,3$.
\end{enumerate}
Applying the calculation method explained in Subsec. \ref{sec:v-num} to the
expectation values
and dispersions, one sees immediately that
\begin{enumerate}[(i)]
  \item the expectation values of $\ln(\hat{v}_a)_{v_1}$ are arithmetic averages
    of the expectation values of the restrictions, and
  \item the dispersions are bounded by an arithmetic average of the dispersions
    of each of the four contributing sectors plus terms of the form
    $\sqrt{\langle \hat{D}_{m}^2 \rangle}$, where the operator $\hat{D}_{m}$
    measures the difference between the total expectation value and that of the
    considered sector, labeled by $m=1,...4$.
\end{enumerate}

On the other hand, in Subsec.
\ref{sec:wdw-limit-disp} we proved that, for each of
the sectors, the relative dispersion of
$\ln(\hat{v}_a)_{v_1}$ ($a=2,3$) is the same in the two
components moving forward/backward in time,
respectively, provided that some mild conditions
(e.g.) on the forward-moving component are satisfied.
In order to extend the result to the ensemble of sectors,
one has to cope with the possible differences between
the expectation values of $\ln(\hat{v}_a)_{v_1}$ mentioned above in point
$(ii)$.

To verify the boundedness of these differences we
observe that the relation between the restrictions of
the eigenfunctions to the subsemilattices
corresponding to the different sectors under
consideration --which is encoded in Eq.~(40) of Ref.
\cite{mmp}--, together with the relation between the
normalization of these restrictions and that of their
WDW limits, imply that the coefficients $A_a$ and
$W_a$ are equal for all the studied sectors [see Eqs.
\eqref{eq:traj-dir-A} and \eqref{eq:traj-dir-W}]. From
the invariance of these coefficients under the
transformation $\ub{\tilde\Phi}_-\to\ub{\tilde\Phi}_+$
and the considerations made in Subsec.
\ref{sec:wdw-limit-disp}, it follows then that the
analyzed differences have indeed a well defined,
finite limit as $v_1\to\infty$.

We also note that, as in the case of the discussion of
Subsec. \ref{sec:wdw-limit-disp}, the above arguments
can be repeated considering the transformation
$\ub{\tilde\Phi}_+\to\ub{\tilde\Phi}_-$, opposite to
the one that we have studied here.

Finally, it is worth commenting that the requirement
that the state have finite expectation values and
dispersions with respect to the operators
$\Omega^{(2)}_a$, although providing a restriction on
the space of states, can be considered quite
reasonable from a physical viewpoint. Actually, since
the choice of $v_1$ as an emergent time (and hence the
role of $\omega_1$ as a ``time frequency'') is
arbitrary, one can exchange the roles of $\omega_1$
and $\omega_2$ and, for a given state represented by
$\tilde{\Phi}(\omega_2,\omega_3)$, consider its
transformation into a different one
\begin{equation}\label{eq:trans1}
\tilde{\Phi}\to\tilde{\Phi}' : \
\tilde{\Phi}'(\omega_2,\omega_3)=\tilde{\Phi}
(\omega_1(\omega_2,\omega_3),\omega_3)
\end{equation}
On the other hand the transformation
\begin{equation}\label{eq:trans2}
\tilde{\Phi}(\omega_2,\omega_3) \mapsto
\frac{\omega_1^2(\omega_2,\omega_3)}{\omega_2^2}
\tilde{\Phi}(\omega_1(\omega_2,\omega_3),\omega_3)
\end{equation}
corresponds just with swapping the coordinate $v_1$
for  $v_2$. Similar transformations can be defined
also for the interchange $v_1 \leftrightarrow v_3$.
Combining both of them one can rewrite the square of
$\Omega^{(2)}_a$ as the square of $\ln|\omega_2|$
acting on the state represented by $\tilde{\Phi}'$.
Therefore, since that square is the main component of
the dispersion, one can relate $\langle\Delta
\Omega^{(2)}_a\rangle$ with the dispersion of the
operator $\ln|\omega_a|$ on $\tilde{\Phi}'$ at least
heuristically. Similar arguments can be applied as
well to $\Sigma^{(2)}_a$.

\section{Effective classical dynamics}\label{app:eff}

The quantum dynamics resulting from the LQC approach
turns out to be surprisingly well reproduced by
certain classical effective dynamics, constructed by
replacing the holonomies and fluxes in the quantum
Hamiltonian constraint by their expectation values
\cite{sv-eff,victor}. This in principle naive
construction has been proven to accurately mimic the
genuine quantum evolution (with discrepancies much
smaller than the quantum dispersions) in several
situations (see e.g.
\cite{aps-imp,b1-szulc,aps-infl}). We sketch here its
derivation for the case of the vacuum Bianchi I model.

We start with the quantum constraint \eqref{C},
applying to it the replacements explained above. The
resulting effective Hamiltonian takes the form
\begin{subequations}\label{eq:effH-def}\begin{align}
H^{\eff} &= -\frac{2}{\gamma^2}
[\Theta^{\eff}_1\Theta^{\eff}_2+\Theta^{\eff}_1
\Theta^{\eff}_3+\Theta^{\eff}_2\Theta^{\eff}_3] , \\
\Theta^{\eff}_i &:= 6\pi\gamma G v_i \sin(b_i) ,
\end{align}\end{subequations}
where the variable $b_i$ ($i=1,2,3$) is the momentum
conjugate to $v_i$ [see Eqs.~\eqref{eq:b-def} and
\eqref{eq:bv-poiss}]. This momentum is compactified as
the unit circle.

Given the Hamiltonian \eqref{eq:effH-def}, one can
easily derive the Hamilton-Jacobi equations for
$\partial_{\tau}v_i$ and $\partial_{\tau}b_i$, where
$\tau$ is a time parameter associated with the
Hamiltonian. Defining the constants of motion
$\Theta^{\eff}_i = 8\pi\gamma G\Kappa_i$ and $\Kappa =
\Kappa_1 + \Kappa_2 + \Kappa_3$ (as in Ref.
\cite{chi2}) we have
\begin{subequations}\begin{align}
\label{veff} (\partial_{\tau} v_i)^2 &= 9 (8\pi G)^4
(\Kappa-\Kappa_i)^2\left[v_i^2 - \left(\frac{4\Kappa_i}{3}
\right)^2\right], \\
\label{b}
\partial_{\tau}b_i &= 3(8\pi G)^2
(\Kappa_i-\Kappa)\sin(b_i).
\end{align}\end{subequations}

Equation \eqref{b} immediately implies that, within
the intervals $b_i\in[0,\pi)$ and $b_i\in[\pi,2\pi)$,
which in the LQC quantization correspond to the
respective subsemilattices
${}^{(4)}\!\lat_{\varepsilon_1}^+$ and
${}^{(4)}\!\lat_{\varepsilon_1+2}^+$ (see Sec.
\ref{sec:lqc-b-rep}), each of the variables $b_i$
provides a good candidate for an internal time. On the
other hand, the general solution to Eq. \eqref{veff}
reads
\begin{align}
v_i(\tau) &= \frac{4}{3}\Kappa_i \cosh[3 (8\pi G)^2
(\Kappa-\Kappa_i)(\tau-\tau_o)],
\end{align}
where $\tau_o$ is a constant representing the time of
the bounce. Therefore, $v_i$ is not monotonous in
time.


\begin{thebibliography}{99}
\bibitem{adv} The invariance can be recovered by
allowing an extra dependence of the
$\bar{\mu}$-parameters of the holonomies on the
(ratios) of the coordinate lengths of the elementary
cell edges, but at the cost of losing an
interpretation of the relation with the full theory of
Loop Quantum Gravity.
\bibitem{F3}
On the contrary, the prescription adopted in Ref.
\cite{awe-b1} leads to the relation
$\bar{\mu}_i\bar{\mu}_j|p_k|=2\Delta$ (such that
$\epsilon_{ijk}\neq0$).
\bibitem{F1}
In what follows, terms in the WDW quantization are
underlined to distinguish them from their analogs in
the loop quantization.
\bibitem{F2}
One must not confuse this Hilbert space with
the space of initial data $\ub{\Hil}_t$.
\bibitem{AltH}
Alternatively, if all the spaces $\ub{\Hil}_t$
coincide, one can build an analog of the operator
$\hat{O}'$ directly on $\ub{\Hil}'$ which then plays
the role of an auxiliary Hilbert space. Such approach
was applied e.g. in Ref. \cite{aps-imp} to construct
$|v|_{\phi}$.
\bibitem{Fn}
The restriction to $L^2(\mathbb{R}^\pm,dv_i)$ can be
identified, e.g., with the restriction to even wave
functions in $L^2(\mathbb{R},dv_i)$, which are mapped
by the Fourier transform to even wave functions in
$L^2(\mathbb{R},db_i)$.
\bibitem{rot-comp}
Note that the rotating components
$e^{\tilde{\varepsilon}_1s}_{\omega_1}$ with support
on ${}^{(4)}\!\lat_{\tilde{\varepsilon}_1}^+$ define
the rotating components with support on
$\lat_{\varepsilon_1}^+$ as
$e^{\varepsilon_1s}_{\omega_1}=
e^{\tilde{\varepsilon}_1s}_{\omega_1} \oplus\,
e^{(\tilde{\varepsilon}_1\pm2)s}_{\omega_1}$, where
the positive sign is taken when
$\tilde{\varepsilon}_1=\varepsilon_1$ and the negative
one if $\tilde{\varepsilon}_1=\varepsilon_1+2$.
\end{thebibliography}
\end{document}